\documentclass[10pt, journal]{IEEEtran}
\IEEEoverridecommandlockouts
\usepackage[noadjust]{cite}
\usepackage{amsmath,amssymb,amsfonts}
\usepackage{algorithmic}
\usepackage{graphicx}
\usepackage{textcomp}
\usepackage{xcolor}

\usepackage{comment}
\usepackage{caption}
\usepackage{subcaption}
\usepackage{multirow}   
\usepackage{multicol}   
\usepackage{url}   
\usepackage{comment}
\usepackage{tabularx}
\usepackage{xcolor,colortbl}
\usepackage{hyperref}
\newcolumntype{x}[1]{%
>{\centering\hspace{0pt}}p{#1}}%

\graphicspath{ {./figs/} }

\def\BibTeX{{\rm B\kern-.05em{\sc i\kern-.025em b}\kern-.08em
    T\kern-.1667em\lower.7ex\hbox{E}\kern-.125emX}}
\begin{document}

\title{Correlation-Aware Neural Networks for DDoS Attack Detection In IoT Systems \\
}

\author{\IEEEauthorblockN{Arvin Hekmati\IEEEauthorrefmark{1}, Nishant Jethwa\IEEEauthorrefmark{1}, Eugenio Grippo\IEEEauthorrefmark{2}, Bhaskar Krishnamachari\IEEEauthorrefmark{1}\IEEEauthorrefmark{2}\\}
\IEEEauthorblockA{\IEEEauthorrefmark{1}Department of Computer Science \\ 
\IEEEauthorrefmark{2}Department of Electrical and Computer Engineering\\ 
University of Southern California\\ Los Angeles, California, USA\\ Email:
  \texttt{\{hekmati, njethwa, egrippo, bkrishna\}@usc.edu}\\[1ex]}}

\maketitle

\begin{abstract} \textbf{
    % Maximum 250 words
    We present a comprehensive study on applying machine learning to detect distributed Denial of service (DDoS) attacks using large-scale Internet of Things (IoT) systems. While prior works and existing DDoS attacks have largely focused on individual nodes transmitting packets at a high volume, we investigate more sophisticated futuristic attacks that use large numbers of IoT devices and camouflage their attack by having each node transmit at a volume typical of benign traffic. We introduce new correlation-aware architectures that take into account the correlation of traffic across IoT nodes, and we also compare the effectiveness of centralized and distributed detection models. We extensively analyze the proposed architectures by evaluating five different neural network models trained on a dataset derived from a 4060-node real-world IoT system. We observe that long short-term memory (LSTM) and a transformer-based model, in conjunction with the architectures that use correlation information of the IoT nodes, provide higher performance (in terms of F1 score and binary accuracy) than the other models and architectures, especially when the attacker camouflages itself by following benign traffic distribution on each transmitting node. For instance, by using the LSTM model, the distributed correlation-aware architecture gives 81\% F1 score for the attacker that camouflages their attack with benign traffic as compared to 35\% for the architecture that does not use correlation information. We also investigate the performance of heuristics for selecting a subset of nodes to share their data for correlation-aware architectures to meet resource constraints.
}
\end{abstract}


\section{Introduction}

    The Internet of things (IoT) has dramatically grown, propelled by the impetuous development of technology \cite{shah2016survey, 7555867}. IoT devices are predicted to be more pervasive in our lives as compared to mobile devices. The number of IoT devices (nodes) connected to the Internet is projected to be around 29 billion devices around the world \cite{numIoTDevices}. Along with the growth of the number of IoT devices, the vulnerability and security risks of IoT devices have also been increasing almost at the same pace. Recent studies by HP security research showed a high average number of vulnerabilities by studying 10 of the most popular IoT devices, such as lack of transport encryption, insecure software/firmware, and cross-site scripting \cite{HPstudy1, HPstudy2}. As a consequence, cybersecurity of IoT devices ought to be developed at an ever-faster rhythm, allowing/accompanying the safe and secure growth of networks \cite{hassan2019current, neshenko2019demystifying}. With this goal in mind, this work addresses one of the most dangerous types of attacks involving IoT systems, namely distributed denial of service (DDoS) attacks  \cite{hallman2017ioddos, kolias2017mirai, bertino2017botnets}. 
    
    In denial of service (DoS) attacks, the attacker tries to disturb the behavior of the victim server in order to block legitimate users' system access by reducing the server's availability. DoS attacks come in many forms, such as user datagram protocol (UDP) flood attacks, where the attacker sends more traffic to the victim server than it is capable of handling \cite{verma2013UDP}, synchronize (SYN) flood attacks, where the attacker sends a transmission control protocol (TCP) connection request to the victim server with spoofed source addresses but never send back the acknowledgment \cite{schuba1997syn}, etc.
    
    In distributed denial of service (DDoS) attacks, attackers get access to many compromised devices, usually called zombies, in order to perform DoS attacks on the victim server. Therefore, the attacker can significantly magnify the effect of the DoS attacks on the victim server\cite{suresh2011ddos}. IoT devices are a perfect choice to perform DDoS attacks given the huge number of IoT devices that have access to the Internet and also their vulnerabilities to become compromised.

    DoS and DDoS attacks could be performed by using the protocols of different layers in the Open Systems Interconnection (OSI) model. DDoS attacks that use the transport layer are the most common type of attacks where the attacker sends as many packets as possible to the victim server through UDP flooding, SYN flooding, etc. We also have other types of DDoS attacks that use the application layer in the OSI model. Attackers try to disturb the behavior of the victim server by tying up its every thread with slow requests. The slow-rate DDoS attacks could be performed by sending data to the victim server at a very slow rate, but fast enough to prevent the connection from getting timed out \cite{Yungaicela2021slowrate}.
    
    One of the most famous IoT-based DDoS attacks is caused by the Mirai botnet. This botnet could bring down the victim servers by infecting thousands of IoT devices and dramatically increase the network traffic directed toward the victim servers on the order of Terabit per second (Tbps), affecting millions of end-users \cite{margolis2017miari, sinanovic2017analysis, marzano2018botnet}. Another example is the Reaper botnet which is a variant of the Mirai botnet and could infect 2.7 million IoT devices \cite{kelley2018reaper}. There are many other botnets that could perform DDoS attacks by using billions of IoT devices around the world and impact many victim servers, and end-users \cite{vishwakarma2020botnet}.

\begin{figure}[h!]
        \centering 
        \begin{subfigure}[]{0.24\textwidth}
            \includegraphics[width=\textwidth]{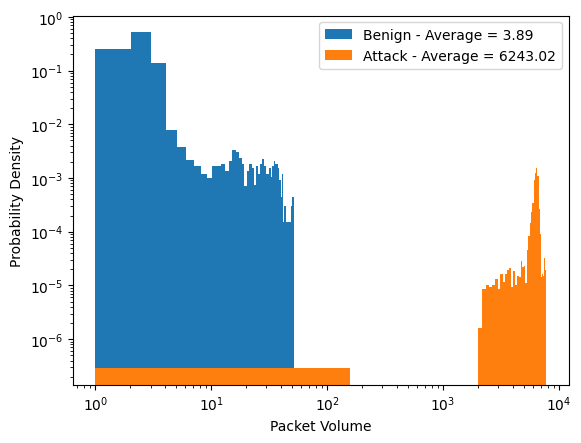}
            \caption{Mirai botnet DDoS attack}
            \label{fig:ddos_packet_volume}
        \end{subfigure}
        \begin{subfigure}[]{0.24\textwidth}
            \includegraphics[width=\textwidth]{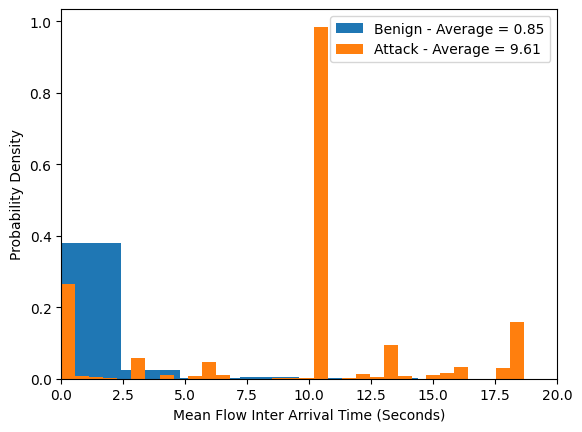}
            \caption{Slow-rate DDoS attack}
            \label{fig:slowrate_ddos_mean_iat}
        \end{subfigure}
        \caption{Benign/attack packet volume (figure left) and mean flow inter-arrival time (figure right) probability density of  real DDoS attacks}
        \label{fig:benign_attack_comparison}
    \end{figure}

    Given this dangerous and dramatically growing threat, we explore the use of machine learning (ML) techniques as the main tool to prevent such attacks \cite{mcdermott2018mirai, sharma2016machine}. One of the most important gaps in the related papers that studies the DDoS detection mechanisms is the fact that all of them consider either the packet volume transmitted from IoT nodes or the packet flow timing while under attack to be significantly (orders of magnitudes) higher as compared to the IoT nodes' benign traffic. For instance, Meidan \textit{et al.} \cite{meidan2018mirai} presented an ML-based technique for detecting the DDoS attacks on a dataset containing the benign and attack traffic of nine IoT nodes. Figure \ref{fig:ddos_packet_volume} shows the probability density of benign and attack traffic packet volume of an IoT node with ID XCS7\_1003 in the Meidan \textit{et al.} \cite{meidan2018mirai} paper. As we can see, the packet volume transmitted from that IoT node while under attack is, on average, 1604 times higher than the packet volume of the IoT node when there is no attack happening. Furthermore, Sharafaldin \textit{et al.} \cite{CICIDS2017} presented a DDoS dataset and also an ML-based technique for detecting DDoS attacks. Their dataset also includes a slow-rate DDoS attack type. Figure \ref{fig:slowrate_ddos_mean_iat} shows the probability density of benign and attack traffic mean flow inter-arrival time (IAT), i.e. the mean time between two packets sent in the flow. As we can see, the average of the mean IAT for the attack traffic is approximately 11 times higher than the benign traffic. These differences between the benign and attack traffic properties could give a huge benefit to the ML models for detecting DDoS attacks. However, in this study, we consider more futuristic DDoS attacks in which attackers can mimic the behavior of the IoT nodes' benign traffic. In fact, given the huge number of IoT devices that are currently available, the attackers could take control of millions of IoT nodes and, camouflage their attack by sending fewer packets from those IoT nodes with similar timing to the benign traffic, thus, disturbing the behavior of the victim server. This futuristic attack which is the focus of this study, is much more sophisticated than traditional attacks where the attacker could expose themselves by either sending many packets or slowing down the data transmission.
    
    We have released an anonymized dataset containing real-trace data from an urban deployment of 4060 IoT devices that records their binary activity~\cite{hekmati2021workshop}. We further enhanced that dataset in our recent work \cite{hekmati2022conference} by adding the packet volume that each IoT device transmits at each timestamp where they are active. The basis for this emulation is grounded in our finding that a real urban IoT benign traffic can be well-modeled as a (truncated) Cauchy distribution~(matching the observation of prior researchers that Ethernet traffic is well modeled by such a distribution~\cite{field2002network}). We have also done a simple analysis of neural network models by only considering one architecture where each IoT node trains individual neural network models without using the correlation information from other nodes \cite{hekmati2022conference}.
    
    In this work, we further enhance the generation of our training dataset, in which we also include the correlation information of IoT nodes' activity in each recorded activity on the dataset. Our proposed “dataset+script” allows the injection of truncated Cauchy distributed attacks, and it also lets the user parametrize the difference between the benign and attack traffic volume. In this way, a diverse training scenario can be effectively generated to improve the training of neural network models. All attack traffic volumes are tuned by a parameter that we call ``$k$” that determines both  the location and scale of the truncated Cauchy distribution for attack traffic volume generation. We further extend our recent work by proposing four architectures to train neural network models for the IoT devices by considering all combinations of either using the correlation information of the IoT devices or not, and either (a) training one central model for all nodes or (b) training distributed individual models for each IoT device. By using the correlation information, each IoT node has access to the packet volume transmitted from other IoT nodes in addition to its packet volume. These architectures are named multiple models with correlation (MM-WC), multiple models without correlation (MM-NC), one model with correlation (OM-WC), and one model without correlation (OM-NC). Furthermore, we considered five different neural network models, namely multi-layered perceptron (MLP), convolutional neural networks (CNN), long short-term memory (LSTM), transformer (TRF), and autoencoders (AEN). We extensively analyzed the performance of proposed architectures by using the mentioned neural network models to decide what is the best architecture/neural network model for detecting DDoS attacks on IoT devices.
    
    Our simulation results indicate that using the correlation information of the nodes significantly helps with detecting DDoS attacks, especially in the case that the attacker is camouflaging the attack by using the benign traffic packet volume distribution. Given the huge number of IoT devices that could potentially be used in performing DDoS attacks, using the correlation information of all nodes results in a massive number of features that neural network models need to learn from. Therefore, we investigated the methods to actively select the nodes for considering the correlation information of the nodes for training the neural network models. Our results indicate that using the Pearson correlation to actively select the nodes for training and prediction purposes, results in a good performance in terms of binary accuracy and F1 score for detecting the DDoS attacks as compared to using the correlation information of all nodes. We make the dataset, attack emulation, and training/testing neural network models and architectures script available as an open-source repository online at \url{https://github.com/ANRGUSC/correlation_aware_ddos_iot}.
    
    The rest of this paper is organized as follows: section \ref{sec:related_works} presents the related works that have been done in this area. In section \ref{sec:original}, we present the raw urban IoT dataset and its benign traffic characteristics; this section also introduces the modeling of the IoT benign traffic (defining the (truncated) Cauchy distribution) and builds the synthetic dataset defining the parameter that will regulate the relation/distance between benign and attack traffic. In section \ref{sec:attack} we illustrate the way we emulate attacks and also the procedure to create the general training dataset to be used for training, validating, and comparing different NN models and architectures to detect DDoS attacks deployed in the customized dataset. Section \ref{sec:defense} presents the four architectures and five neural network models we used for this analysis. In section \ref{sec:iot_constrained}, we discuss new methods for considering the correlation information of IoT nodes while considering their constrained resources by actively selecting IoT nodes for sharing their information in correlation-aware architectures. Section \ref{sec:results} presents the evaluation and analysis of the introduced models. Lastly, section \ref{sec:conclusion} summarizes this work and proposes future research steps. 
    
\section{Related Works}
\label{sec:related_works}
    Prior works have explored various machine learning models for detecting DDoS attacks. Most of the papers in this area considered the transport layer DDoS attacks, where attackers send as many packets as possible to the victim server to disturb its behavior. Doshi \textit{et al.} \cite{doshi2018ddosML} generated the training data by setting up an environment with three IoT devices and simulated the Mirai Botnet attack. They have developed various classification models such as random forests, K-nearest neighbors, support vector machines, and neural networks to be run on network middleboxes such as routers, firewalls, or network switches to detect DDoS attacks, and concluded that almost all of the models could achieve 0.99 accuracy. Chen \textit{et al.} \cite{chen2020sdnML} used 9 IoT devices at a university campus to collect the data through IoT gateways and transmit the data to cloud servers via the software-defined network (SDN) switches. Then, they implemented a decision tree to run on the IoT gateways and SDN controllers in order to detect and block the DDoS attacks and achieved an F1 score of 0.99. Similarly, Mohammed \textit{et al.} \cite{mohammed2018sdnML} implemented a naive Bayes classification technique based on the NSL-KDD dataset on a central server and then tested their model using the traffic information captured from four SDN controllers and achieved an F1 score of 0.98.
    Syed \textit{et al.} \cite{syed2020brokerML} proposed an application layer DoS attack detection framework that uses a machine learning-based method on the Message Queuing Telemetry Transport (MQTT) brokers. They tested their framework by using the average one-dependence estimator (AODE), C4.5 decision trees, and multi-layer perceptron (MLP) machine learning models on three virtual machines and achieved 0.99 accuracy. Meidan \textit{et al.}~\cite{meidan2018mirai} collected their training data by infecting nine commercial IoT devices and proposed N-BaIoT anomaly detection method that uses autoencoders to classify the benign and attacked IoT traffic with 0.99 accuracy. %Their proposed method runs only on IoT devices and does not consider the correlation information that is embedded in the IoT devices' traffic while they are under attack. However, in this paper, we propose two architectures that consider the correlation information in the IoT nodes traffic in order to detect the DDoS attack. 

    Here we also mention some of the works that studied slow-rate DDoS detection mechanisms. Yungaicela-Naula \textit{et al.} \cite{Yungaicela2021slowrate} used CICDDoS 2017 \cite{CICIDS2017}, and CICDDoS 2019 \cite{UNB_two} and utilized machine learning and deep learning methods such as support vector machine (SVM), K nearest neighbor (KNN), multi-layer perception (MLP), etc. to detect DDoS attacks. Their experiments showed that they could achieve up to 98\% accuracy for the traditional DDoS attack and up to 95\% accuracy for slow-rate DDoS attacks. Similarly, Nugraha and Murthy \cite{Nugraha2020slowrate} proposed a hybrid CNN and LSTM neural network model that could achieve up to 99\% accuracy on a dataset that they self-developed for slow-rate DDoS attacks. Muraleedharan and Janet \cite{Muraleedharan2021slowrate} used a random forest (RF) classifier to detect slow-rate DDoS attacks, and their model could achieve up to \%99 accuracy. Cheng \textit{et al.} \cite{Cheng2020slowrate} studied the performance of random forest, K nearest neighbor, naive Bayes, and support vector machine on a self-developed dataset for slow-rate DDoS attack detection on the network switches and controllers. Their analysis showed that ML models could achieve up to \%99 accuracy. There are many other works that studied the DDoS detection on IoT devices using machine learning methods some of them are presented in table \ref{table:ml-iot}.
    
    In all the related works discussed above, we are observing that the datasets and features that they have used to do DDoS attack classification have very different properties for the attack and benign traffic, where attack traffic volume/timing is orders of magnitude higher than the benign traffic volume/timing. However, in this work, we use the tunable parameter $k$ that can tune the packet volume being transmitted from IoT devices with regular timing as compared to the benign traffic and help the attacker to camouflage the attack in the benign traffic. Furthermore, most of the mentioned related works studied either a central neural network model to run on the cloud servers, SDN switches, etc., or a few of them studied the distributed neural network model to run on the IoT devices and did not provide a comparison for the distributed versus central architectures. However, in this work, we extensively analyzed both of these scenarios and compared their performance against each other. Furthermore, none of the papers mentioned above considered the correlation information of IoT nodes with relation to each other while designing the DDoS detection mechanisms. However, in this paper, we also study the architectures that are correlation-aware and consider the correlation information of the IoT nodes for detection purposes. Moreover, our paper proposes and uses a large-scale IoT DDoS dataset with more than 4000 nodes, and we introduce a training dataset that considers correlation information of the IoT nodes against other papers which use datasets that have a limited number of IoT nodes without having the correlation information of the nodes included. Table \ref{table:datasets} presents an overview of datasets in this field with the respective number of IoT nodes used in creating the dataset.

    \begin{table*}[]
    \centering
    \caption{Selected papers with ML based methods for detecting DDoS attacks on IoT}
    \label{table:ml-iot}
    \begin{tabular}{lllll}
    \hline
    \textbf{Reference} &
      \textbf{Dataset} &
      \textbf{Detection Method} &
      \textbf{Centralized/Distributed} &
      \textbf{Inference Device} \\ \hline
    Doshi \textit{et al.} \cite{doshi2018ddosML} &
      Self-developed &
      RF, KNN, SVM, MLP &
      Centralized &
      Network middle box \\
    Chen \textit{et al.} \cite{chen2020sdnML} &
      Self-developed &
      DT &
      Centralized &
      SDN controller \\
    Syed \textit{et al.} \cite{syed2020brokerML} &
      Self-developed &
      MLP, AODE, DT &
      Centralized &
      MQTT broker \\
    Meidan \textit{et al.} \cite{meidan2018mirai} &
      N-BaIoT \cite{meidan2018mirai} &
      AEN &
      Distributed &
      IoT node \\
    Liu \textit{et al.} \cite{liu2018sdnML} &
      Self-developed &
      RL &
      Centralized &
      SDN controller \\
    Roopak \textit{et al.} \cite{roopak2019mlIoT} &
      Self-developed &
      MLP, CNN, LSTM &
      Centralized &
      Unclear \\
    Gurulakshmi \textit{et al.} \cite{gurulakshmi2018mlIoT} &
      Self-developed &
      SVM &
      Centralized &
      Unclear \\
    Mohammed \textit{et al.} \cite{mohammed2018sdnML} &
      NSL-KDD \cite{nslkddDataset} &
      NB &
      Centralized &
      SDN controller \\
    Zekri \textit{et al.} \cite{zekri2017mlcloud} &
      Self-developed &
      DT &
      Centralized &
      Cloud \\
    Blaise \textit{et al.} \cite{Blaise2020MLIoT} &
      CTU-13 \cite{ctu13dataset} &
      MLP, SVM, RF, LR &
      Centralized &
      Unclear \\
    Soe \textit{et al.} \cite{Soe2020MLIoT} &
      N-BaIoT \cite{meidan2018mirai} &
      MLP, DT, NB &
      Centralized &
      Unclear \\
    Nugraha \textit{et al.} \cite{Nugraha2020MLIoT} &
      CTU-13 \cite{ctu13dataset} &
      MLP, CNN, LSTM &
      Centralized &
      Unclear \\
    Kumar \textit{et al.} \cite{Kumar2019MLIoT} &
      Self-developed &
      RF, KNN, GNB &
      Centralized &
      IoT gateway \\ 
    Yungaicela \textit{et al.} \cite{Yungaicela2021slowrate} &
      CICDDoS \cite{CICIDS2017},\cite{UNB_two} &
      RF, SVM, KNN, MLP, CNN, GRU, LSTM &
      Centralized &
      Server \\ 
    Cheng \textit{et al.} \cite{Cheng2020slowrate} &
      Self-developed &
      SVM, NP, KNN, RF &
      Centralized &
      Controller and Switch \\ 
      \hline 
    \multicolumn{5}{c}{\begin{tabular}[c]{@{}c@{}}Where MLP: Multilayer Perceptron, CNN: Convolutional Neural Network, LSTM: Long Short Term Memory,\\ AEN: Autoencoder, RL: Reinforcement Learning, DT: Decision Tree, RF: Random Forest, KNN: K Nearest Neighbor,\\ SVM: Support Vector Machine, AODE: Average One-Dependence Estimator, NB: Naive Bayes,\\ GNB: Gaussian Naive Bayes, LR: Logistic Regression, GRU: Gated Recurrent Unit \end{tabular}}
    \end{tabular}
    \end{table*}

    \begin{table*}[]
    \centering
    \caption{Related Papers with IoT datasets}
    %\begin{tabular}{|l|c|c|c|c|c|c|}
    \begin{tabularx}{\textwidth}{Xlcccccc}
    
    \hline
    %\multicolumn{1}{|c|}{\textbf{DATASET}} & \textbf{Date} & \textbf{Number of Nodes} & \textbf{IoT specific/General} & \textbf{Binary activity or Traffic Volume} & \textbf{Benign/Attack traffic}\\ \hline
    \textbf{Reference} & \textbf{Date} & \textbf{Number of Nodes} & \textbf{IoT specific/General} & \textbf{Binary activity or Traffic Volume} & \textbf{Benign/Attack traffic}\\ \hline

    DARPA 2000\cite{DARPA_2000} & 2000 & 60 & general  & traffic volume & both\\
    
    CAIDA UCSD DDoS Attack 2007 \cite{CAIDA_UCSD} & 2007 & unclear & general & traffic volume & attack\\

    Shiravi et. al. \cite{shiravi2012toward} &  2012 & 24 & general & traffic volume &    both              \\ 
    
    %Singh et. al. \cite{singh2015approach}   & 2015  & 1000 & general & traffic volume &                \\ 
        
    CICDDoS 2017 \cite{CICIDS2017}  &  2017  & 25 & general & traffic volume &    both                \\ 
    
    Meidan et. al. \cite{meidan2018mirai} & 2018 & 9 & IoT specific & traffic volume & both\\ 
    
    CSE-CIC-IDS2018 on AWS \cite{UNB_three}  & 2018 & 450 & general & traffic volume & attack\\ 
        
    CICDDoS 2019 \cite{UNB_two}   & 2019   & 25 & general & traffic volume &      attack              \\ 
    
    The Bot-IoT Dataset (Univ. of NSW) \cite{UNSW_two} & 2019 & unclear & IoT specific & traffic volume & both\\

    Ullah et. al \cite{bot_iot} & 2020 & 42 & IoT specific & traffic volume & both \\  

    Erhan et. al. \cite{erhan2020bougazicci}. &  2020 & 4000 & general &  traffic volume & both\\ 
    Hekmati et. al. \cite{hekmati2021workshop}. &  2021 & 4060 & IoT specific & binary activity & both\\ 
    Hekmati et. al. \cite{hekmati2022conference} &  2021 & 4060 & IoT specific & traffic volume & both\\ \hline
     
    \end{tabularx}
    \label{table:datasets}    
    \end{table*}

%\input{aux_notes_1_data_processing}
%\vspace{30mm}
\section{Original and Benign Activity Datasets}
\label{sec:original}
    The original (raw) data of the binary activity of the IoT nodes have been collected from the IoT nodes that were deployed in an urban area. \footnote{The source of this data, originally presented in~\cite{hekmati2021workshop}, has been anonymized for privacy and security reasons.}. The features provided in the raw dataset are the node ID, the latitude and longitude location of the IoT node, and the binary activity status of the IoT devices. A record of the IoT node appears on the raw dataset whenever there is an activity status change for an IoT node, i.e., after each change in the activity status of the IoT nodes, a new record will be added to the raw dataset. The raw dataset has 4060 nodes with one month's worth of data with no missing data points. 
    
    As mentioned before, we have activity status changes of the nodes in the raw dataset. This means that the nodes that have more changes in their activity will have a higher number of records in the raw dataset. Therefore, using the raw dataset for the purpose of training neural network models will provide a bias toward the behavior and information of the nodes that have the highest number of activity changes.  In order to address this issue, we provided a script that takes the raw dataset with the record of activity status changes of the IoT nodes and also a time step, called $t_s$, and generates a new dataset called benign dataset that presents the activity status of each IoT node every $t_s$ seconds. Therefore, the activity status of each IoT node will be present in the benign dataset every $t_s$ seconds, which helps the neural network models to learn the activity behavior of all IoT nodes at all time stamps instead of just learning the behavior of the IoT nodes that have the highest changes in their activity status. Furthermore, the python script can also get a beginning and ending date for generating the benign dataset in the case the IoT node has constrained resources to train neural network models on all the benign dataset.
    
    As mentioned above, the benign dataset will only have the binary activity status of the IoT nodes. In order to enhance the dataset, we also add the packet volume that gets transmitted from the active IoT nodes at each time step. Meidan et al. \cite{meidan2018mirai}  studied the DDoS attack detection on 9 IoT nodes in the real world and presented a dataset that contains the packet volume transmitted from each IoT node every 10 seconds both in the case that nodes are under attack and also the case that nodes are not under attack. We used the data related to a security camera IoT node with ID XCS7\_1003 in the Meidan et al. \cite{meidan2018mirai} paper as a source for generating the benign/attack packet volume of the IoT nodes in our dataset. In the literature, various distributions have been used to estimate the packet volume of network traffic \cite{chandrasekaran2009survey}. We analyzed 80 different distributions for estimating the packet volume of the IoT nodes based on the real information that we had for the security camera datatset that was provided in \cite{meidan2018mirai}. Among the 80 different distributions that we analyzed, the Cauchy distribution fitted best to the packet volume activity of the security camera node in terms of having the minimum square error. Since the Cauchy distribution has an unusual nature due to being unbounded in value and not having a defined mean, we use a truncated Cauchy distribution instead with a low of 0, and a high of maximum packet volume observed in the real data of the security camera IoT node. Our analysis in finding truncated Cauchy distribution as the best distribution for estimating the packet volume of the IoT nodes is also compatible with the previous finding for modeling network traffic~\cite{field2002network}. In order to generate the packet volume in the benign dataset, we will generate the packet volume transmitted from the IoT node whenever the node is active based on the fitted truncated Cauchy distribution. On the other hand, when the node is not active, the packet volume transmission from the IoT node will be zero. Note that in this paper, we are presenting an event-driven IoT dataset which means that the IoT nodes will transmit packets whenever they are active. Table \ref{table:data-sample} presents a few sample data points in the benign dataset.\\
    
    \begin{table}[h!]
    \caption{Sample Data Points in Benign Dataset}
    \centering
    \begin{tabular}{|c|c|c|c|c|c|}
    \hline
    \textbf{NODE} & \textbf{LAT} & \textbf{LNG} & \textbf{TIME}                                                 & \textbf{ACTIVE} & \textbf{PACKET} \\ \hline
    1          & 33       & 40       & \begin{tabular}[c]{@{}c@{}}2021-01-01\\ 23:00:00\end{tabular} & 0               & 0               \\ \hline
    1          & 33       & 40       & \begin{tabular}[c]{@{}c@{}}2021-01-01\\ 23:10:00\end{tabular} & 0               & 0               \\ \hline
    1          & 33       & 40       & \begin{tabular}[c]{@{}c@{}}2021-01-01\\ 23:20:00\end{tabular} & 1               & 9               \\ \hline
    1          & 33       & 40       & \begin{tabular}[c]{@{}c@{}}2021-01-01\\ 23:30:00\end{tabular} & 1               & 11              \\ \hline
    \end{tabular}
    \label{table:data-sample}
    \end{table}

    In order to generate the packet volume of the IoT nodes when they go under attack, we define a new truncated Cauchy distribution with the following parameters: 

    \begin{align}
        \label{eqn:k_x0}
        x_a &= (1+k) \cdot x_b \\
        \label{eqn:k_gamma}
        \gamma_a &= (1+k) \cdot \gamma_b \\
        \label{eqn:k_max}
        m_a &= (1+k) \cdot m_b
    \end{align}
    
    where, $x_b$, $\gamma_b$, $m_b$ refer to the location, scale, and maximum packet volume of the truncated Cauchy distribution of the benign traffic, respectively, while $x_a$, $\gamma_a$ and $m_a$ are the location, scale, and maximum packet volume of the generated truncated Cauchy distribution of the attack traffic, respectively. $k$ is the tunable parameter for generating packet volumes with higher location, scale, and maximum packet volume as compared to benign traffic where $k>=0$. Note that when the DDoS attack gets performed by the attacker, the IoT nodes will become active and transmit packets, whether they were active or not previously. As we mentioned before, the related works in this area are only considering the cases where in the DDoS attack, either IoT nodes are sending as many packets as possible or slowing down the data transmission to the victim server in order to disturb its behavior. However, in this work, we are considering the case that the attacker can perfectly camouflage the attack with benign traffic. More specifically, for low $k$ values, the attacker is trying to camouflage with background traffic since it is sending packets by using the benign traffic pattern of the IoT nodes, which is harder to detect the DDoS attack. On the other hand, when the attacker uses a higher value for $k$, larger packet volumes will be transmitted to the victim server, but it will also be easier to get detected. While one could potentially use three different parameters for creating new $x_a$, $\gamma_a$, and $m_a$, we use only one parameter $k$ just for simplicity. 
        
\begin{figure}[h!]
        \centering 
        \begin{subfigure}[]{0.24\textwidth}
            \includegraphics[width=\textwidth]{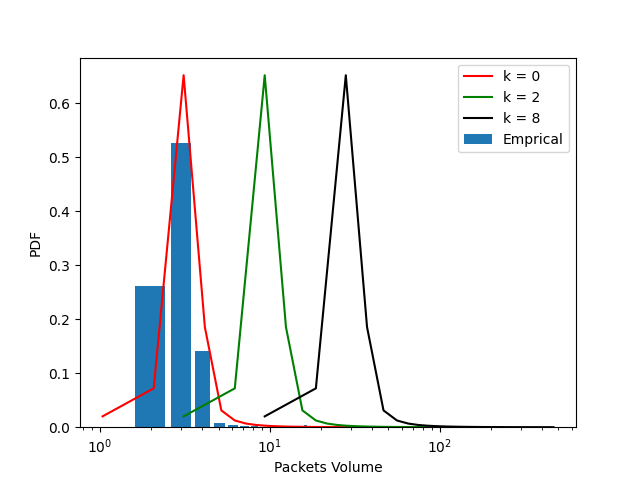}
            \caption{PDF}
            \label{fig:packet_pdf}
        \end{subfigure}
        \begin{subfigure}[]{0.24\textwidth}
            \includegraphics[width=\textwidth]{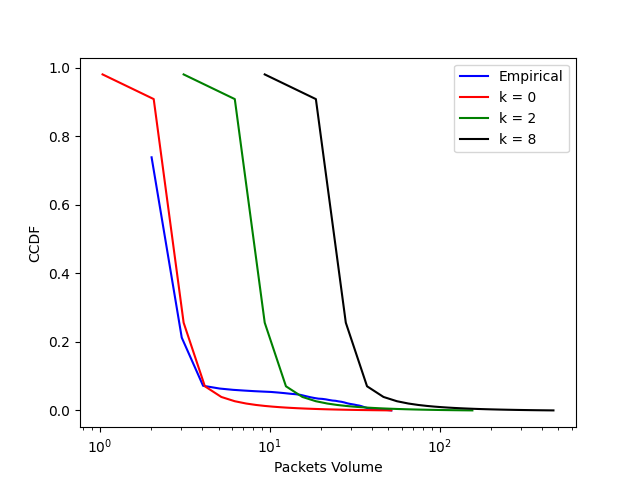}
            \caption{CCDF}
            \label{fig:packet_ccdf}
        \end{subfigure}
        \caption{Real packet volume distribution vs truncated Cauchy distribution}
        \label{fig:packet_distribution}
    \end{figure}

    In order to better understand the behavior of the tunable parameter $k$ and also compare the truncated Cauchy distribution estimation against the real packet volume distribution, here we present figures \ref{fig:packet_pdf} and \ref{fig:packet_ccdf} that show the probability density function (PDF) and complementary cumulative distribution function (CCDF) for the real empirical benign traffic data in a blue color besides the truncated Cauchy distribution with $k$ values of 0, 2, and 8. By assigning the $k$ value to 0, the packet volume generated by the truncated Cauchy distribution should be similar to the real packet volume distribution. As we can see in figures \ref{fig:packet_pdf} and \ref{fig:packet_ccdf}, the truncated Cauchy distribution with $k = 0$ (red color) is well fitted to the real empirical packet volume distribution (color blue). By increasing the value of the tunable parameter, $k$, we are basically increasing the location, scale, and maximum packet volume of the truncated Cauchy distribution, which results in having higher probabilities for larger packet volumes. Using high values for $k$ results in transferring huge amounts of packets during the attack. In this scenario, the attacker can disturb the behavior of the victim server with fewer number of IoT nodes. However, using lower values of $k$, although lower packet volumes will be generated by each IoT node during the attack, the attacker could use a large number of IoT nodes to transmit packets to the victim server and disturb its behavior. 
    
\section{Attack Mechanism}
\label{sec:attack}
    
    This section presents how synthetic DDoS attacks are generated by using the IoT nodes, given the packet volume distribution presented in section \ref{sec:original}. Furthermore, the procedure for generating the general training dataset is also discussed in this section.

    \subsection{Generating DDoS attack}
    \label{sec:attack-attack_dataset}
        In this paper, we synthetically generate a DDoS attack on the IoT nodes by setting all nodes which are under attack to active status for the duration of the attack and assigning what packet volume to be transmitted at each time slot. Recall that since we have an event-driven IoT dataset, when the nodes go under attack, they will start transmitting packets, whether they have been active or not previously. In order to generate the packet volume transmitted while the nodes are under attack, we use the equations (\ref{eqn:k_x0}), (\ref{eqn:k_gamma}), and (\ref{eqn:k_max}) to define the truncated Cauchy distribution for a given tunable parameter $k$ and sample in i.i.d fashion from that distribution. The DDoS attack performed by the attacker can have four different parameters to be set: the start time of the attack ($a_s$), the duration of the attack ($a_d$), the ratio of the nodes to be used by the attacker ($a_r$), and the tunable attack packet distribution parameter ($k$). In this way, we can synthesize various desired attacks that an attacker could potentially perform. Therefore, the neural network models will be able to learn the behavior of the nodes in different attack scenarios and predict what IoT nodes are under attack at what time.

    \subsection{Generating General Training Dataset}
    \label{sec:attack-training_dataset}
        Here we present the procedure for creating the \textit{general} training dataset that contains all features that may be used by different architectures proposed in section \ref{sec:defense-architectures}. By performing the attack on the benign dataset, we will generate a labeled general training dataset for supervised machine learning training. At each sample of the general training dataset for a specific IoT node $i$, we store the node ID, the time stamp, and also the packet volume that is being transmitted through that node. In addition to the packet volume of node $i$ at each sample of the training dataset, we also store the packet volume for all other nodes in the general training dataset. This means that at each time stamp of the general training dataset, we will have access to the packet volume that is being transmitted through node $i$ and all other nodes. Furthermore, in order to distinguish the records for each node, we also add one-hot encoding to the general training dataset. Finally, we will have the label of each sample that shows whether that sample represents the node being under attack or not. Note that having the record of the packet volumes that are being transmitted from all nodes at each sample of the general training dataset will help to understand the correlation behavior of the nodes and will be used in training the correlation-aware architectures. As mentioned before, we will propose four different architectures that either use the correlation information and one-hot encoding in training or not. This matter will be elaborated more in section \ref{sec:defense-architectures}. Table \ref{table:general_training_dataset} represents a sample training dataset containing the information of only two nodes where N\_1 and N\_2 represent nodes 1 and 2, respectively, and P\_1 and P\_2 represent the packet volume of nodes 1 and 2, respectively. Note that, in order to detect the DDoS attack at each timestamp, we also need the information of the previous time slots to understand the behavior of the node through time. Therefore, the training dataset could be considered a time-series dataset. Thus, we stack the past $n_t$ entries of each sample in the training dataset, to predict the attack status of that specific sample. In the python script provided, one could also tune the $n_t$ based on the resources available for the IoT devices. Having higher values for $n_t$ would help the neural network model to better understand the behavior of the IoT devices, but it also requires higher computation resources for training the neural network models.
        
        \begin{table}[]
        \caption{Sample Data Points in General Training Dataset}
        \centering
        \begin{tabular}{|c|c|c|c|c|c|c|}
        \hline
        \textbf{Node} & \textbf{Time} & \textbf{N\_1} & \textbf{N\_2} & \textbf{P\_1} & \textbf{P\_2} & \textbf{Attacked} \\ \hline
        1 & \begin{tabular}[c]{@{}c@{}}2021-01-01\\ 23:20:00\end{tabular} & 1 & 0 & 12 & 9  & 0 \\ \hline
        1 & \begin{tabular}[c]{@{}c@{}}2021-01-01\\ 23:30:00\end{tabular} & 1 & 0 & 13 & 8  & 1 \\ \hline
        2 & \begin{tabular}[c]{@{}c@{}}2021-01-01\\ 23:20:00\end{tabular} & 0 & 1 & 9  & 12 & 0 \\ \hline
        2 & \begin{tabular}[c]{@{}c@{}}2021-01-01\\ 23:30:00\end{tabular} & 0 & 1 & 8  & 13 & 1 \\ \hline
        \end{tabular}
        \label{table:general_training_dataset}
        \end{table}

\section{Defense Mechanism}
\label{sec:defense}
     In this section, we present the architectures and neural network models that we used for detecting DDoS attacks. The performance comparison of these architectures and neural network models are presented in section \ref{sec:results}.
     
     \subsection{Architectures}
     \label{sec:defense-architectures}
         
         We introduce four different architectures to train the neural network models based on whether to use the correlation information of nodes' activity or not and whether to train distributed individual models for each IoT node or train one central model for all nodes. Here we present the details of each architecture:
         
         \begin{itemize}
             \item \textbf{Multiple models without correlation (MM-NC)}: In this architecture, we train individual neural network models for each IoT node. Therefore, each node will have one model trained just for itself. Furthermore, we do not use the correlation information of nodes' activity, i.e., we only use the packet volume of each node through time in order to train the models and do not include the packet volume of other nodes in the training dataset. Since we are training individual models for each node, we will also not use one-hot encoding features from the general training dataset in this architecture. 
             \item \textbf{Multiple models with correlation (MM-WC)}: In this architecture, we train individual neural network models for each IoT node. Therefore, each node will have one model trained just for itself. Furthermore, we use the correlation information of nodes' activity, i.e., in addition to each node's packet volume, we also use the packet volume of other nodes to capture the correlation information of nodes' activity in training. Since we are training individual models for each node, we will also not use one-hot encoding features from the general training dataset in this architecture. 
             \item \textbf{One model without correlation (OM-NC)}: In this architecture, we train one neural network model for all IoT nodes. Therefore, there will be a central node/server that trains the neural network model, and all IoT nodes will use that model for detecting DDoS attacks. Furthermore, we do not use the correlation information of nodes' activity, i.e., we only use the packet volume of each node through time in order to train the models. Since we are training one central model for all nodes, in order to distinguish the information of each node from other nodes, we use one-hot encoding from the general training dataset.
             \item \textbf{One model with correlation (OM-WC)}: In this architecture, we train one neural network model for all IoT nodes. Therefore, there will be a central node/server that trains the neural network model, and all IoT nodes will use that model for detecting DDoS attacks. Furthermore, we use the correlation information of nodes' activity, i.e., in addition to each node's packet volume, we also use the packet volume of other nodes to capture the correlation information of nodes' activity in training. Since we are training one model for all nodes, in order to distinguish the information of each node from other nodes, we use one-hot encoding from the general training dataset.
        \end{itemize}
        
        Using the correlation information of the nodes in the MM-WC and OM-WC architectures could potentially help the neural network models to better predict the DDoS attack since attackers usually use many IoT devices for performing DDoS attacks. Using one model for training the neural network models in the OM-WC and OM-NC architecture will potentially learn the general behavior of the nodes' activity and inactivity better while the MM-WC and MM-NC architectures that use individual models for training neural network models for each node could better personalize the model for each IoT node based on their behavior.
    
    \subsection{Neural Network Models}
    \label{sec:defense-neural_network_models}
        For each architecture discussed in section \ref{sec:defense-architectures}, we use five different neural network models to do binary classification for detecting the DDoS attack on the IoT nodes:
    
        \begin{itemize}
            \item \textbf{Multilayer Perceptron (MLP)}: This is the simplest model of the feed-forward artificial neural network that consists of one input layer, one output layer, and one or more hidden layers \cite{Pal1992MLP}. In this paper, the input layer is followed by one dense layer with 5 neurons and Rectified Linear Unit (ReLU) activation. In the MM-WC and OM-WC architectures, the dense layer also has a 30\% dropout rate, while in the MM-NC and OM-NC architectures, the dense layer does not perform dropout. The output is a single neuron with the Sigmoid activation function.
            
            \item \textbf{Convolutional Neural Network (CNN)}: CNNs are similar to MLP models but use mathematical convolution operation in at least one of their hidden layers instead of general matrix multiplication \cite{Albawi2017CNN}. In this paper, the model's input layer is followed by one 1-dimensional convolution layer with 5 filters, kernel size of 3, and ReLU activation function. In the MM-WC and OM-WC architectures, the CNN layer also has a 30\% dropout rate, while in the MM-NC and OM-NC architectures, the CNN layer does not have a dropout rate. The CNN layer is followed by a 1-dimensional max-pooling layer with a pool size of 2, and then a flattened layer. Finally, the output is a single neuron with the Sigmoid activation function. 
            
            \item \textbf{Long Short-Term Memory (LSTM)}: This is a neural network model that can process a sequence of data and is a suitable choice for time-series datasets \cite{Hochreiter1997LSTM}. In this paper, the model's input layer is followed by one LSTM layer with 4 units and a hyperbolic tangent activation function. In the MM-WC and OM-WC architectures, the LSTM layer also has an L2 regularization factor of 0.3, while in the MM-NC and OM-NC architectures, the LSTM layer does not have an L2 regularization. Finally, the output is a single neuron with the Sigmoid activation function. 
            
            \item \textbf{Transformer (TRF)}: Transformer models are a neural network model that solely relies on the attention mechanism to calculate dependencies between the input and output \cite{Vaswani2017TRF}. This model has an encoder and decoder, but for the purpose of binary classification, we will only need the encoder part.  In this model, the input layer is followed by a multi-head attention layer with one attention head, and the size of each attention head for query and key is 1. In the MM-WC and OM-WC architectures, the multi-head attention layer also has an L2 regularization factor of 0.3, while the MM-NC, and OM-NC architectures, do not have L2 regularization. The multi-head attention layer is followed by the global average pooling layer. Finally, the output is a single neuron with the Sigmoid activation function.

            \item \textbf{Autoencoder (AEN)}:  Autoencoders are a type of neural network model that consists of two parts. The first part tries to encode the input into a compressed representation and the second part tries to decode the compressed representation into the original input \cite{Bank2020AEN}. In fact, this model tries to learn a meaningful compressed representation of the input. In this paper, the input layer is followed by an encoder that consists of five dense layers with 256, 128, 64, 32, and 16 neurons with a ReLU activation function, each followed by batch normalization. The decoder consists of five dense layers with 16, 32, 64, 128, and 256 neurons with a ReLU activation function, each followed by batch normalization. Finally, the output dimension will be the same as the input layer. For the binary classification, the encoder is followed by a dense layer with 8 neurons and a ReLU activation function, and the output will be a single neuron with a Sigmoid activation function. By using the AEN model, we essentially train the encoder to encode our input dataset into a latent space. Then, we design a classification model that gets this latent space as input and predicts whether the latent features show malicious behavior or not.
        \end{itemize}
        \begin{figure*}
            \centering
            \includegraphics[width=0.99\textwidth]{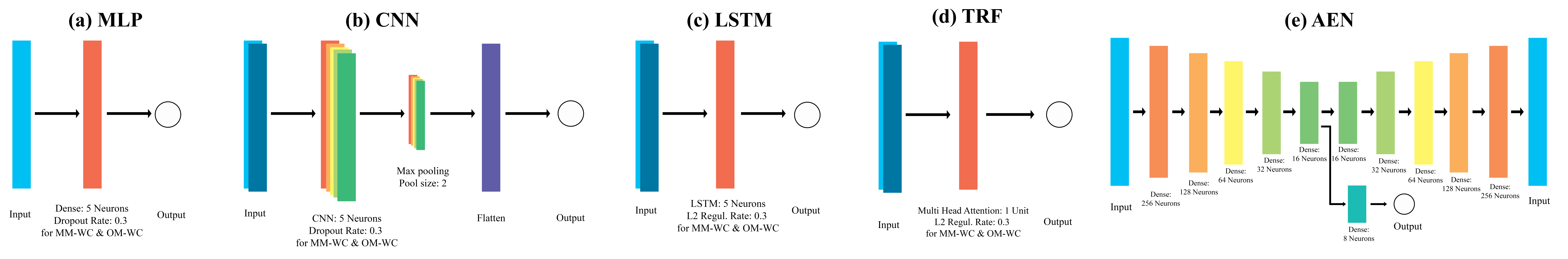}
            \caption{Neural Network Models}
            \label{fig:nn_models}
        \end{figure*}
        
        Figure \ref{fig:nn_models} also presents the architectures discussed above. Given that the correlation-aware architectures have access to the correlation information of the IoT nodes, i.e. MM-WC and OM-WC, we used regularization techniques such as dropout and L2 regularization methods to address the overfitting issues that we observed while training the neural network models. This is worth mentioning that the parameters and hyperparameters for designing/training the neural network model has been selected through a grid search that provides the highest performance. Furthermore, since we are dealing with an unbalanced dataset to do binary classification, we should weigh more the minority class in training the neural network models and also intelligently set the initial bias of the layers. We used the practical method presented in \cite{tensorflow_unbalanced_dataset} to set the class weights and also initial bias. The following formulation is used to define the weights for each class and also the initial bias of the layers:
        
        \begin{align}
            \label{eqn:w_n}
            w_n &= \frac{pos + neg}{2\ neg} \\
            \label{eqn:w_p}
            w_p &= \frac{pos + neg}{2\ pos} \\
            \label{eqn:b_0}
            b_0 &= \log(\frac{pos}{neg})
        \end{align}
        
        where, $w_n$ and $w_p$ represents the weight for the negative(not attacked) and positive(attacked) class, respectively. $b_0$ represents the initial bias of the layers. $neg$ and $pos$ represent the total number of negative and positive samples, respectively. 
        
\section{IoT Nodes with Constrained Resources}
\label{sec:iot_constrained}
\begin{figure}[h!]
        \centering 
        \begin{subfigure}[]{0.24\textwidth}
            \includegraphics[width=\textwidth]{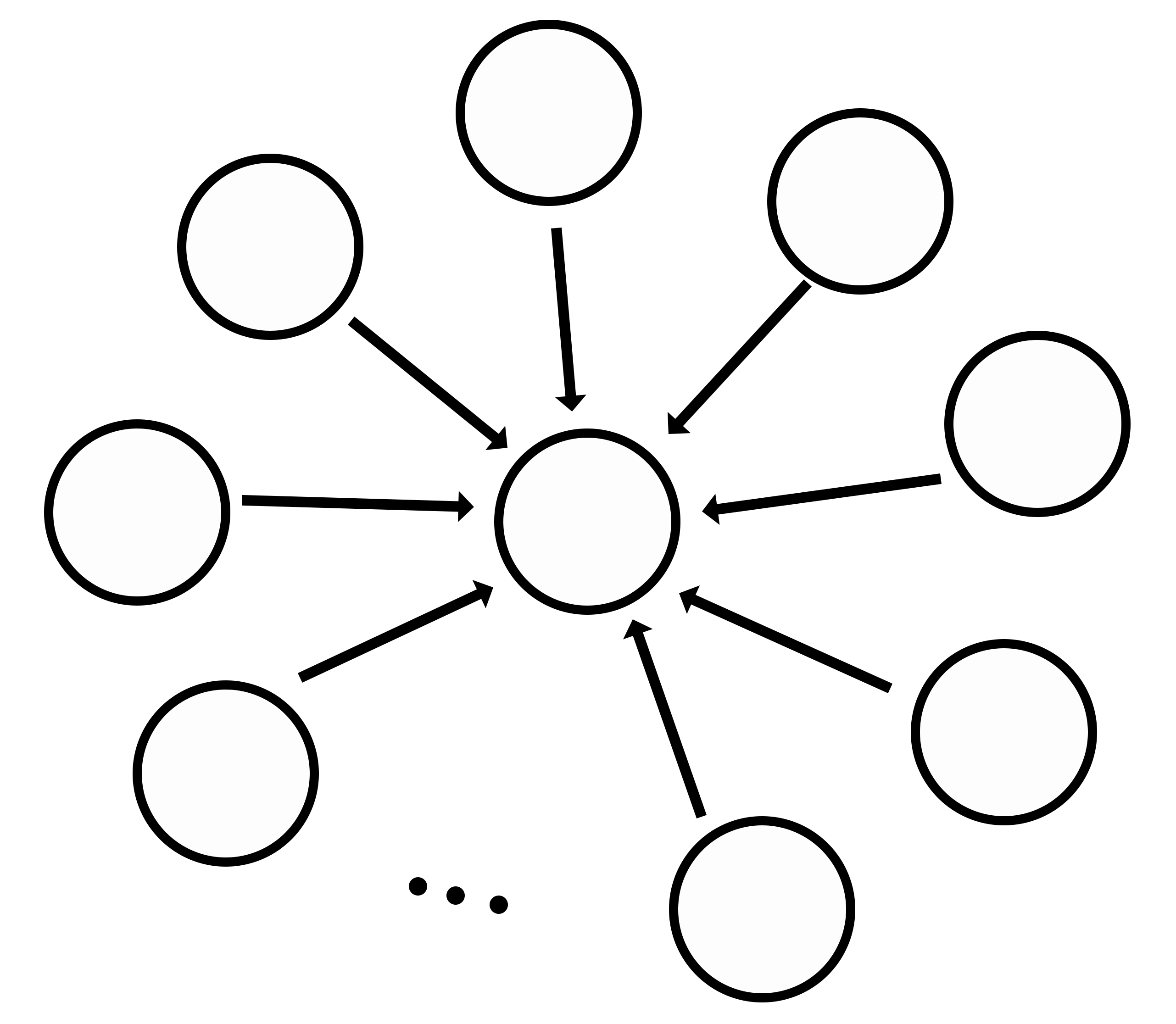}
            \caption{All Nodes}
            \label{fig:constrained_iot_all_nodes}
        \end{subfigure}
        \begin{subfigure}[]{0.24\textwidth}
            \includegraphics[width=\textwidth]{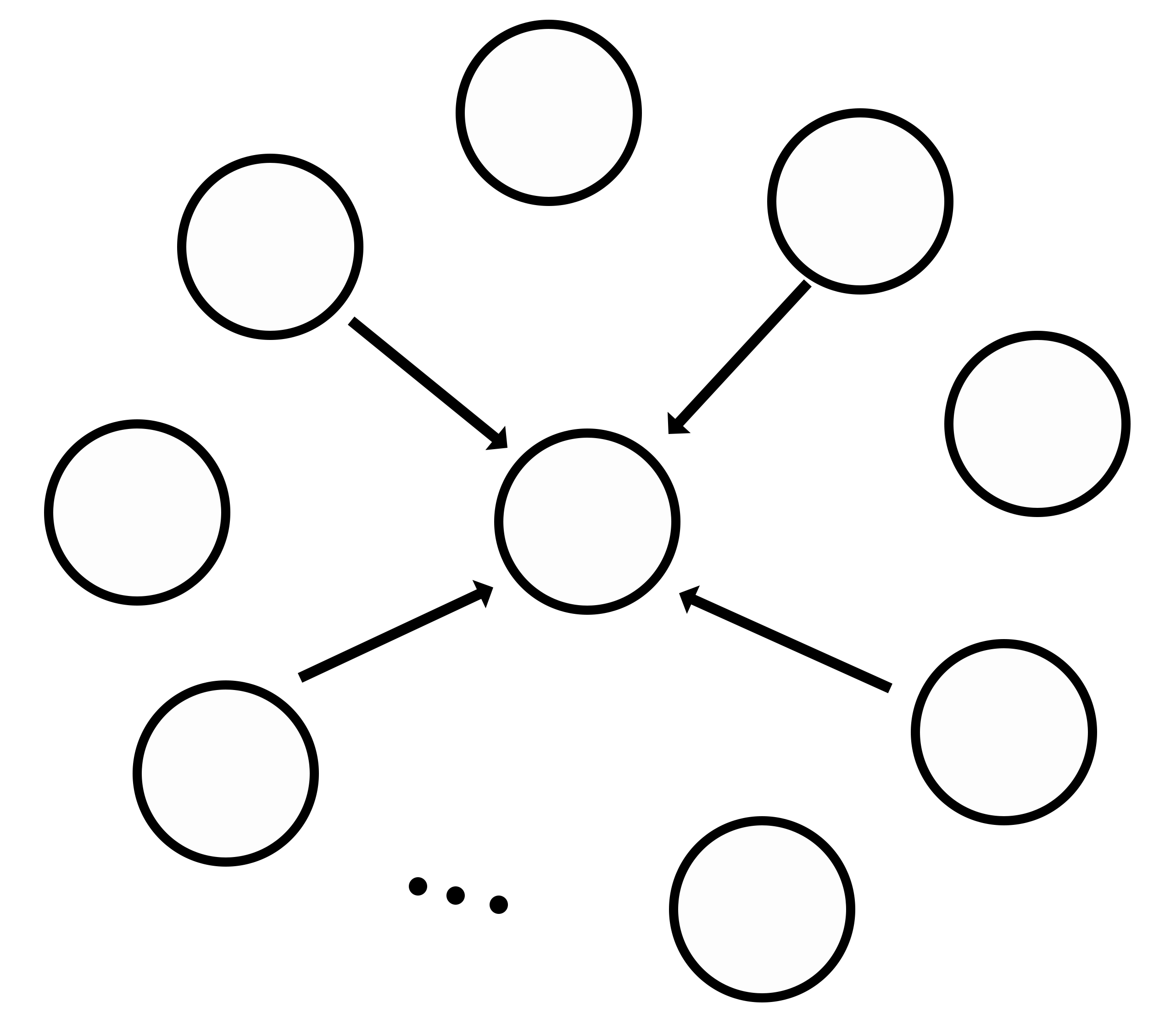}
            \caption{Selected Nodes}
            \label{fig:constrained_iot_selected_nodes}
        \end{subfigure}
        \caption{Using all nodes vs selected nodes for sharing information in correlation-aware architectures}
        \label{fig:constrained_iot}
    \end{figure}

    The architectures introduced in section \ref{sec:defense-architectures} that use correlation information of the nodes, i.e. MM-WC and OM-WC, need to combine the information of \textit{all} nodes in order to train the neural network models. In fact, the packet volume of \textit{all} of the nodes will be used as an input to the neural network models. Using the correlation-aware architectures, results in a massive number of input features to be fed to the neural network models and make it infeasible to use in real-world scenarios with a huge number of IoT devices that have constrained resources and computation power. In order to address this issue, we propose that each IoT node uses a fraction of the other IoT nodes' information as input to their neural network models. As we can see in figure \ref{fig:constrained_iot}, we could either use all of the nodes' traffic information (figure \ref{fig:constrained_iot_all_nodes}) or use selected nodes' information (figure \ref{fig:constrained_iot_selected_nodes}) to be shared with a specific node for the purpose of training neural networks and predicting DDoS attacks. By actively selecting nodes, we can hugely save time and computation in training neural networks and predicting DDoS attacks on IoT nodes with constrained resources.
    
    In this section, we provide the following five heuristic solutions to select important IoT nodes for sharing their information with a specific IoT node in order to train the neural network models in the case of using architectures that utilize the correlation information:
    \begin{itemize}
        \item \textbf{All nodes}: In this method, we use the packet volume of all nodes for training the neural network models. This is the default method that has been discussed in previous sections, and we expect to have the best performance due to using the information of all nodes at the cost of needing high computation power.
        \item \textbf{Pearson correlation}: In this method, we use Pearson correlation to calculate the correlation of the behavior of the nodes' activity based on the packet volume that they are transmitting throughout the day. More precisely, for each node $i$, we calculate the Pearson correlation of node $i$ with all other nodes and find the top $n$ nodes that have the highest correlation with the node $i$. Finally, in training the correlation-aware architectures for each node, we only use the information of top $n$ nodes that have the highest Pearson correlation with node $i$, in addition to the information of node $i$.
        \item \textbf{SHAP}: In this method, we use SHapley Additive exPlanations (SHAP) \cite{Lundberg2017SHAP} to determine which features are the most important in making a decision of whether a node is under attack or not. More precisely, after training a neural network model for node $i$ by using the information of all nodes, we run SHAP method to determine what are the top $n$ features in making the decision for node $i$. Then, we will train a new neural network model that only uses the information of the top $n$ features in making the decision. Note that, in this method, we always need to train the neural network model on all of the nodes first and then determine what nodes provide the most information for the detection by using SHAP. Then, we need to retrain the neural network models by using the top $n$ features. This is not a practical solution to be used in the real world, but we also implemented this method to compare it against other solutions.
        \item \textbf{Nearest Neighbor}: In this method, we use the euclidean distance of the nodes to calculate the correlation of the behavior of the nodes' activity based on their distance. More precisely, for each node $i$, we calculate the euclidean distance of node $i$ with all other nodes and find the top $n$ nodes that are closest to the node $i$. Finally, in training the correlation-aware architectures for each node, we only use the information of the top closest $n$ nodes in addition to the information of node $i$.
        \item \textbf{Random}: In this method, in order to train the correlation-aware architectures for node $i$, we randomly select $n$ other nodes and use their information in addition to the node's $i$ information in order to train the neural network models. Then, we will repeat this process for the desired number of times and calculate the average of the evaluation metrics of all runs. The random method gives us a baseline to compare it against the other solutions mentioned above.
    \end{itemize}

\section{Neural Network Models and Architectures Evaluation}
    \label{sec:results}
    In this section, we evaluate the proposed neural network models and also architectures by doing extensive simulations to analyze each architecture/model's performance in different scenarios.
    
    \subsection{Experiment Setup}
    
    In our simulations, we used a time step of 10 minutes, i.e. $t_s = 600 seconds$, for creating the benign dataset by using the method mentioned in section \ref{sec:original}, but one could use other desired time steps by using the script provided. The introduced enhanced dataset has 4060 IoT nodes. In this subsection, we randomly selected 50 nodes to do the simulations. Here we present the activity analysis of the full dataset with 4060 nodes against the 50 random nodes that we have chosen from the dataset for the purpose of evaluating neural network models.
    
\begin{figure}
        \centering
        \includegraphics[width=\columnwidth]{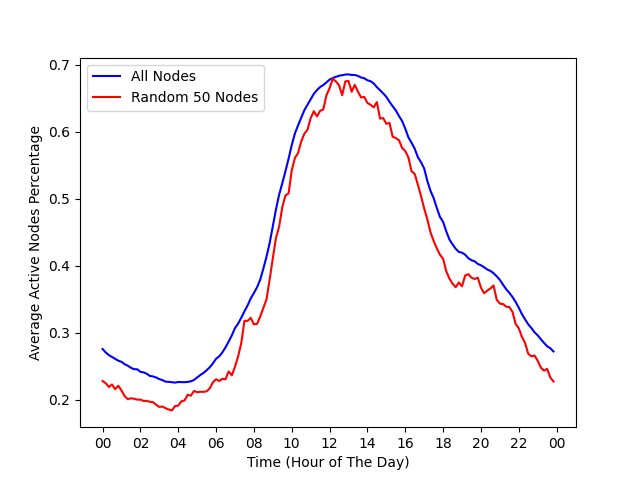}
        \caption{Active Nodes Percentage vs Time}
        \label{fig:active_nodes_percentage}
    \end{figure}

    Figure \ref{fig:active_nodes_percentage} presents the mean number of active nodes in the benign dataset versus the time of the day by considering all nodes and also the 50 randomly selected nodes. As we can see, up to 70\% of the nodes, get activated around the middle of the day, but by midnight only about 20\% of the nodes are active. We can also see that the 50 randomly selected IoT nodes have similar activity behavior as compared to the activity behavior of all nodes.
    
\begin{figure}[]
        \centering 
        \begin{subfigure}[]{0.24\textwidth}
            \includegraphics[width=\textwidth]{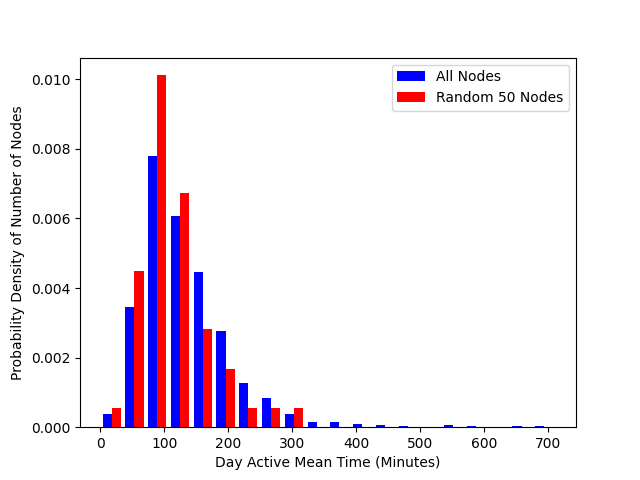}
            \caption{PDF of nodes mean activity in the day}
            \label{fig:day_active}
        \end{subfigure}
        \begin{subfigure}[]{0.24\textwidth}
            \includegraphics[width=\textwidth]{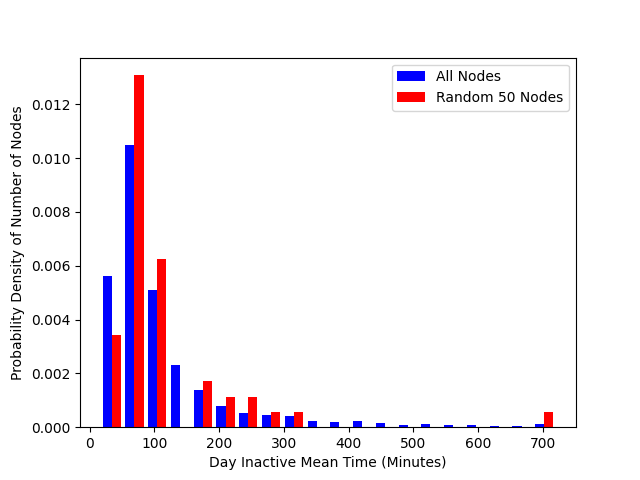}
            \caption{PDF of nodes mean inactivity in the day}
            \label{fig:day_inactive}
        \end{subfigure}
        \begin{subfigure}[]{0.24\textwidth}
            \includegraphics[width=\textwidth]{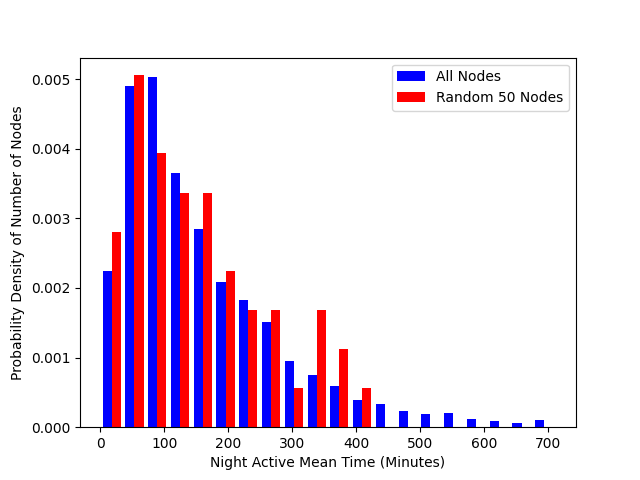}
            \caption{PDF of nodes mean activity in the night}
            \label{fig:night_active}
        \end{subfigure}
        \begin{subfigure}[]{0.24\textwidth}
            \includegraphics[width=\textwidth]{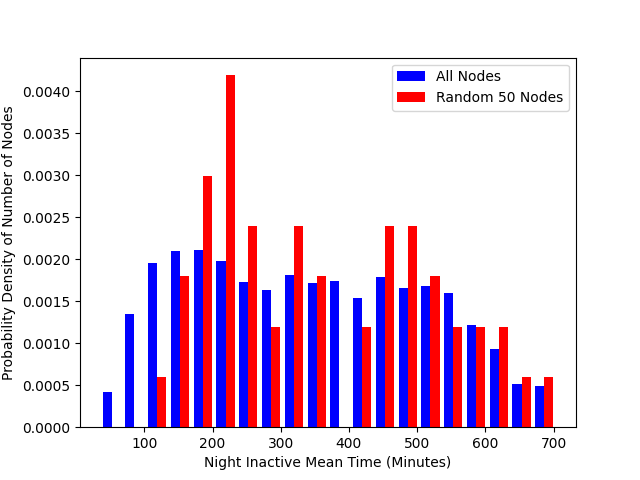}
            \caption{PDF of nodes mean inactivity in the night}
            \label{fig:night_inactive}
        \end{subfigure}
        \caption{Probability density function (PDF) of nodes mean activity/inactivity for day and night}
        \label{fig:nodes_mean_activity}
    \end{figure}

    Figure \ref{fig:nodes_mean_activity} shows the probability density function of nodes' mean activity/inactivity throughout the day (from 8 AM to 8 PM) and night (from 8 PM to 8 AM) for all nodes and also the 50 randomly selected nodes. We observe that nodes are primarily active for around 100 minutes and also mostly inactive for around 80 minutes during the day. During the night, nodes are primarily active for around 80 minutes, but they are mostly inactive for around 220 minutes. Another takeaway from figure \ref{fig:nodes_mean_activity} is that the activity and inactivity behavior of the 50 randomly selected nodes in the day and night is also similar to the behavior of all nodes. Therefore based on our observation from figures \ref{fig:active_nodes_percentage} and \ref{fig:nodes_mean_activity}, the 50 randomly selected IoT nodes closely represent the behavior of all nodes in the dataset, and we can safely extend our simulation findings by using the results that we get by training the neural network models and architectures over the 50 randomly selected nodes.
    
    The whole dataset has one month's worth of data, and in this simulation, we used 4, 1, and 3 different days of the dataset for the training, validation, and testing of the neural network models, respectively. As we discussed in section \ref{sec:attack-attack_dataset}, we can synthesize the DDoS attacks by providing the attack properties to the python script. The properties of the DDoS attacks that we used in this simulation are the following:
    \begin{itemize}
        \item The attacks can start at 2 AM, 6 AM, and 12 PM. We chose different start times for the attacks to make sure the models get trained to predict the DDoS attacks even at the high trafficked times of the day like at noon and do not have a bias towards the time of the day that the DDoS attacks may start.
        \item The duration of the attacks could be 4 hours, 8 hours, or 16 hours. We used different hours of attacks to make sure the model could predict both short and long duration DDoS attacks.
        \item The ratio of the nodes that go under attack could be selected from 0.5 and 1. We chose different attack ratios due to the fact that sometimes the attackers do not use all the compromised IoT devices to perform the DDoS attack.
        \item Attack parameter, $k$, which tunes the packet volume that is being transmitted from the IoT nodes that go under attack, could be chosen from 0,  0.1, 0.3, 0.5, 0.7, and 1. When $k$ is close to 0, the attacker is using the benign traffic distribution to transmit packets to the victim server, which makes it hard to get detected. On the other hand, when $k$ is close to 1, the packets transmitted from the attacked nodes will be higher, and therefore more harm is caused to the victim server but potentially easier to get detected.
    \end{itemize}
    
    Note that, in order to generate the attack and general training dataset, we use the combination of all possible attack scenarios by using the values of the attack properties discussed above in order to consider all possible combinations for training the neural network models.
    
    In order to generate the training dataset, we considered a time window of 10, i.e. $n_t=10$, which means the neural network models will make predictions based on the information on the past 10 time slots of each sample. After generating the training datasets, it has been shuffled to not have prediction bias towards the time of the day. All proposed neural network models have been trained for 3 epochs using 32 batches.

    \subsection{Binary Classification Metrics}
    \label{sec:results-binary_classification_metrics}
        In this subsection, we present the performance of the different architectures and neural network models in terms of their binary accuracy, F1 score, and area under curve (AUC) versus the tunable parameter, $k$, over the testing dataset. Furthermore, we also present the receiver operating characteristic (ROC) curve for $k=0$ to compare the performance of the different architectures and different neural network models for the hardest scenario of detecting DDoS attacks while attackers use benign traffic distribution to generate packets. Note that, in order to calculate the mentioned performance metrics for each value of $k$, we are calculating the average of the model's prediction performance over all other attack properties, such as attack start time, duration, and the ratio of the nodes that go under attack. By these figures, we can effectively compare the performance of different architectures and neural network models over each value of $k$ and conclude what architecture/model performs the best for each scenario.
    
\begin{figure}[h]
            \centering 
            \begin{subfigure}[]{0.24\textwidth}
                \includegraphics[width=\textwidth]{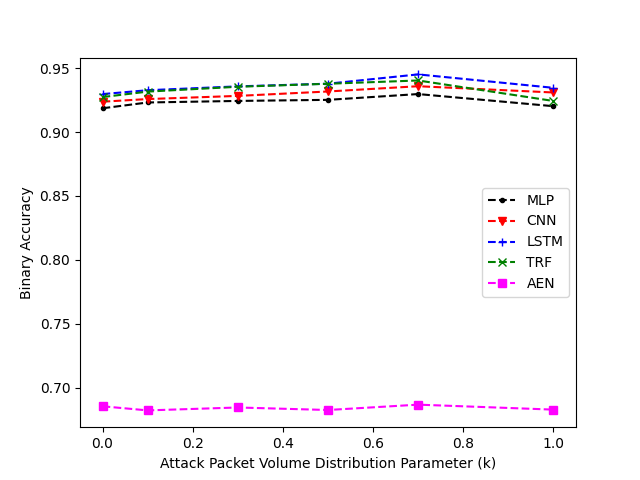}
                \caption{Binary Accuracy}
                \label{fig:mm_wc_binary_accuracy}
            \end{subfigure}
            \begin{subfigure}[]{0.24\textwidth}
                \includegraphics[width=\textwidth]{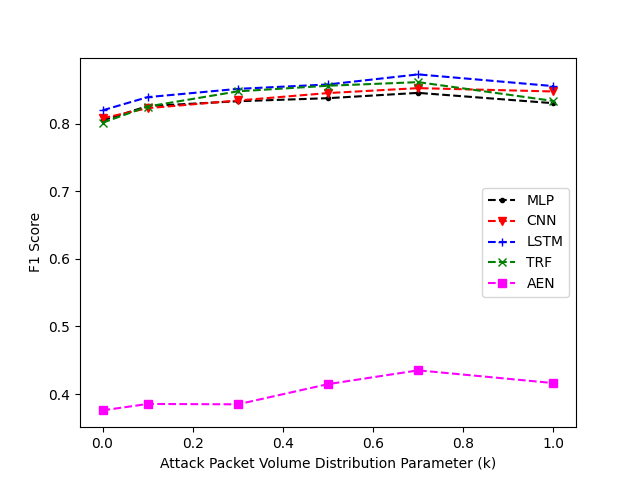}
                \caption{F1 Score}
                \label{fig:mm_wc_f1_score}
            \end{subfigure}
            \begin{subfigure}[]{0.24\textwidth}
                \includegraphics[width=\textwidth]{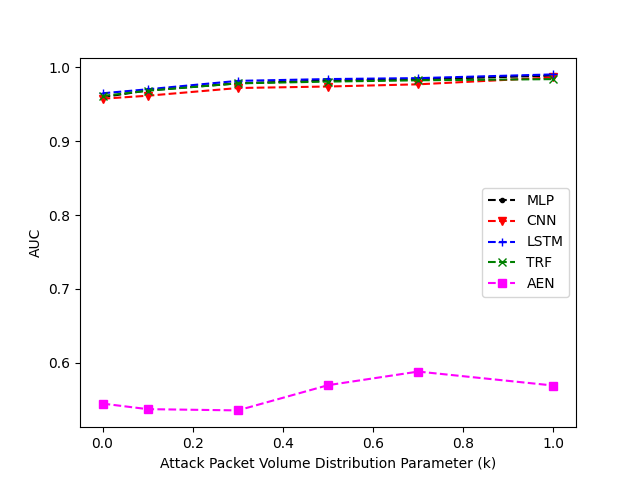}
                \caption{AUC}
                \label{fig:mm_wc_auc}
            \end{subfigure}
            \begin{subfigure}[]{0.24\textwidth}
                \includegraphics[width=\textwidth]{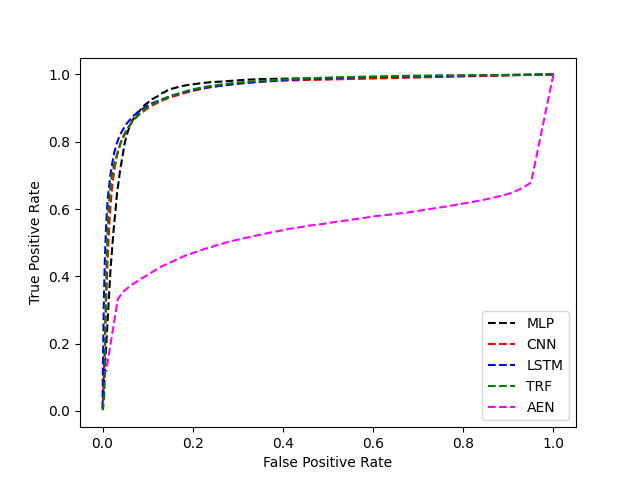}
                \caption{ROC Curve - $k = 0$}
                \label{fig:mm_wc_roc_k_0}
            \end{subfigure}
            \caption{Compare different neural network models' performance by using the multiple models with correlation (MM-WC) architecture}
            \label{fig:mm_wc_performance}
        \end{figure}

        Figure \ref{fig:mm_wc_performance} presents the performance of the multiple models with correlation architecture, i.e. MM-WC, across different neural network models discussed in section \ref{sec:defense-neural_network_models}. As we can see, LSTM neural network model has superior performance as compared to other models almost in all metrics, with an F1 score of between 0.82 and 0.87. TRF has a slightly lower performance as compared to the LSTM model. Following them, CNN and MLP are performing best, with an F1 score between 0.8 and 0.83. AEN has the worst performance for the MM-WC architecture, with an F1 score between 0.39 and 0.45. This shows us that AEN can not learn the behavior of the benign traffic due to the many input features of the MM-WC architecture. This matter is elaborated more at the end of this subsection.
    
\begin{figure}[h]
            \centering 
            \begin{subfigure}[]{0.24\textwidth}
                \includegraphics[width=\textwidth]{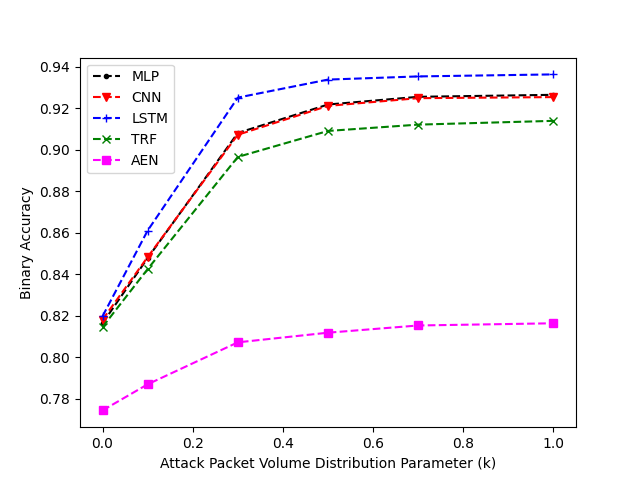}
                \caption{Binary Accuracy}
                \label{fig:mm_nc_binary_accuracy}
            \end{subfigure}
            \begin{subfigure}[]{0.24\textwidth}
                \includegraphics[width=\textwidth]{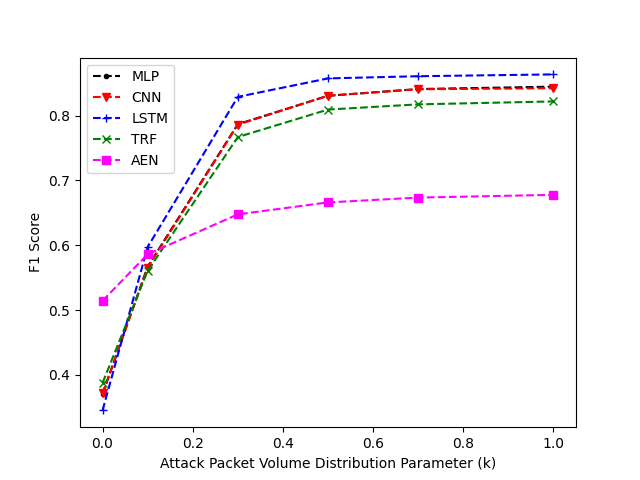}
                \caption{F1 Score}
                \label{fig:mm_nc_f1_score}
            \end{subfigure}
            \begin{subfigure}[]{0.24\textwidth}
                \includegraphics[width=\textwidth]{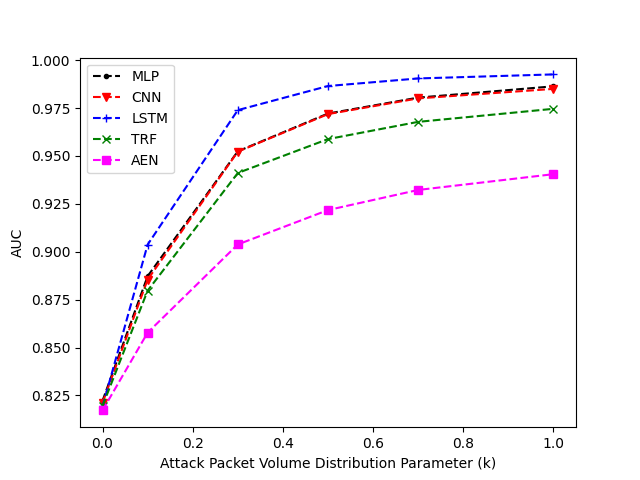}
                \caption{AUC}
                \label{fig:mm_nc_auc}
            \end{subfigure}
            \begin{subfigure}[]{0.24\textwidth}
                \includegraphics[width=\textwidth]{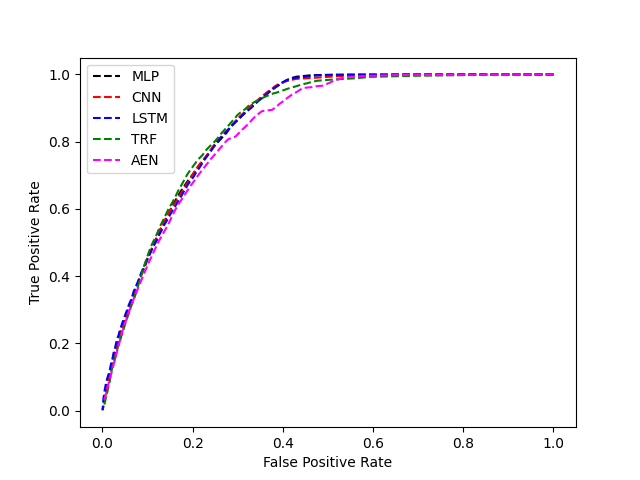}
                \caption{ROC Curve - $k = 0$}
                \label{fig:mm_nc_roc_k_0}
            \end{subfigure}
            \caption{Compare different neural network models' performance by using the multiple models without correlation (MM-NC) architecture}
            \label{fig:mm_nc_performance}
        \end{figure}

        Figure \ref{fig:mm_nc_performance} presents the performance of the multiple models without correlation architecture, i.e. MM-NC, across different neural network models. As we can see, the performance of all neural network models is increasing with higher values of $k$. LSTM neural network model has superior performance as compared to other models with an F1 score between 0.35 and 0.85. Following the LSTM model, CNN, MLP, and TRF are doing best with similar performance. Again, AEN has the worst performance for the MM-NC architecture as compared to other models with an F1 score between 0.51 and 0.65. One thing to note is that AEN has a better performance by using MM-NC architecture as compared to the MM-WC architecture. Furthermore, we are observing that AEN is showing superior performance for the case $k=0$ as compared to other models for the F1 score metric but doing similar or worse for other metrics, i.e. binary accuracy, AUC, and ROC curve. The improved performance of the AEN model in the MM-NC architecture as compared to the MM-WC architecture shows that fewer input features are more suitable for the AEN architecture that can better learn the behavior of the benign traffic and compare it to the attack traffic for the DDoS attack detection.

\begin{figure}[h]
            \centering 
            \begin{subfigure}[]{0.24\textwidth}
                \includegraphics[width=\textwidth]{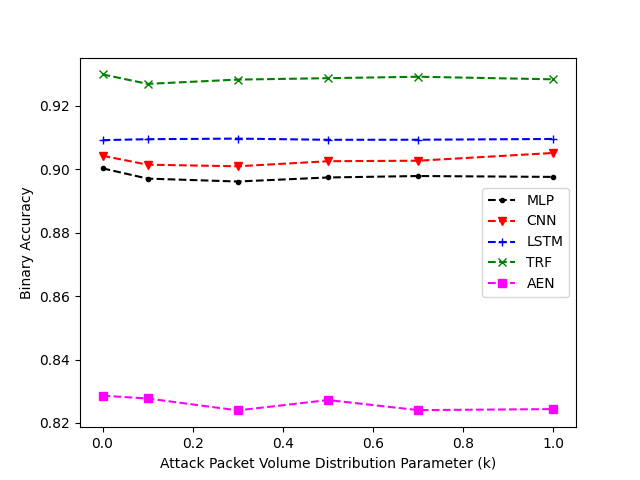}
                \caption{Binary Accuracy}
                \label{fig:om_wc_binary_accuracy}
            \end{subfigure}
            \begin{subfigure}[]{0.24\textwidth}
                \includegraphics[width=\textwidth]{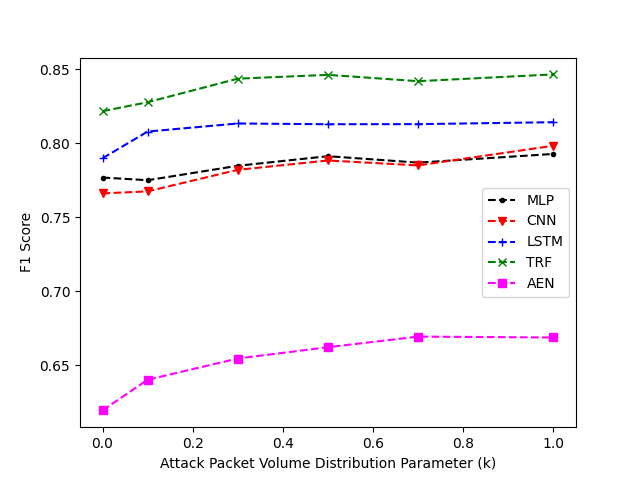}
                \caption{F1 Score}
                \label{fig:om_wc_f1_score}
            \end{subfigure}
            \begin{subfigure}[]{0.24\textwidth}
                \includegraphics[width=\textwidth]{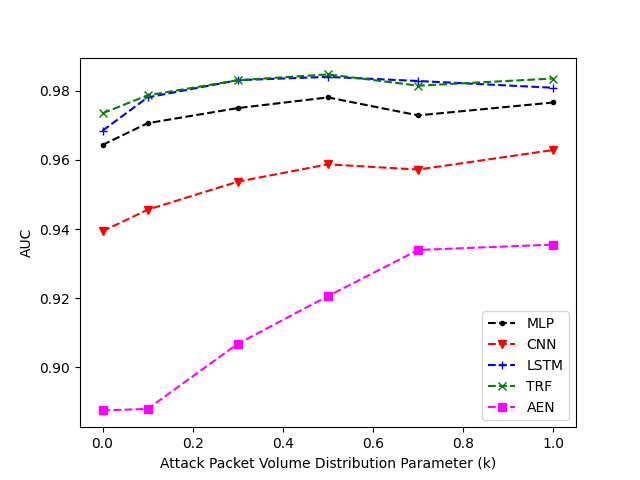}
                \caption{AUC}
                \label{fig:om_wc_auc}
            \end{subfigure}
            \begin{subfigure}[]{0.24\textwidth}
                \includegraphics[width=\textwidth]{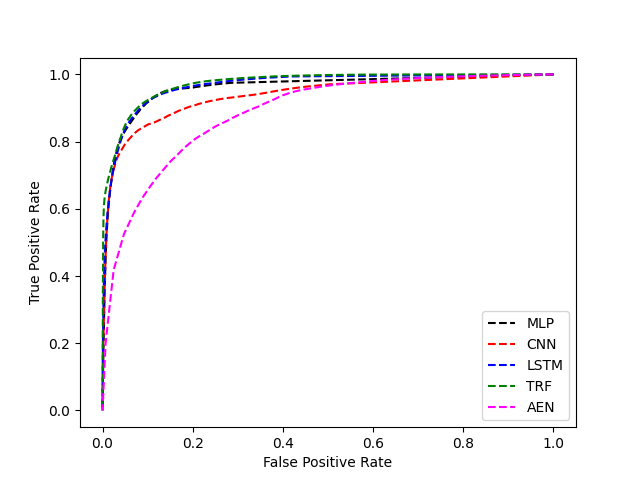}
                \caption{ROC Curve - $k = 0$}
                \label{fig:om_wc_roc_k_0}
            \end{subfigure}
            \caption{Compare different neural network models' performance by using the one model with correlation (OM-WC) architecture}
            \label{fig:om_wc_performance}
        \end{figure}

        Figure \ref{fig:om_wc_performance} presents the performance of the one model with correlation architecture, i.e. OM-WC, across different neural network models. As we can see, TRF neural network model has superior performance as compared to other models, with an F1 score between 0.82 and 0.85. Following that, the LSTM model is performing best, with an F1 score between 0.79 and 0.81. Finally, MLP and CNN are showing similar performance with an F1 score between 0.76 and 0.78. Again, AEN has the worst performance for the OM-WC architecture. This figure, similar to the MM-WC architecture, supports the fact that AEN could not handle many input features to the model. We can also observe that the TRF model can handle more complex architectures, such as OM-WC architecture, where we have many input features to the neural network model as compared to other models.

\begin{figure}[h]
            \centering 
            \begin{subfigure}[]{0.24\textwidth}
                \includegraphics[width=\textwidth]{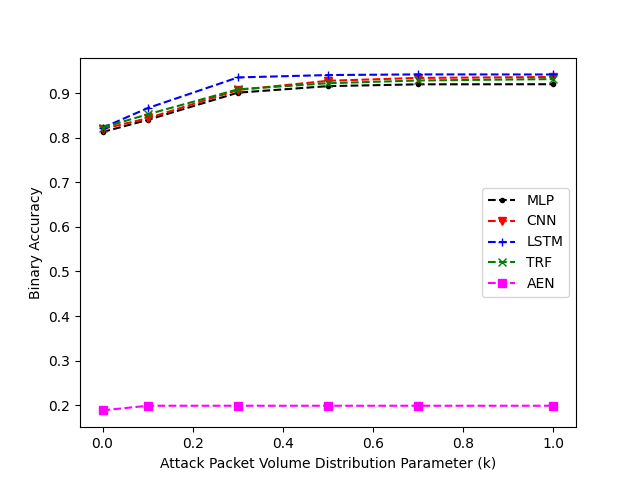}
                \caption{Binary Accuracy}
                \label{fig:om_nc_binary_accuracy}
            \end{subfigure}
            \begin{subfigure}[]{0.24\textwidth}
                \includegraphics[width=\textwidth]{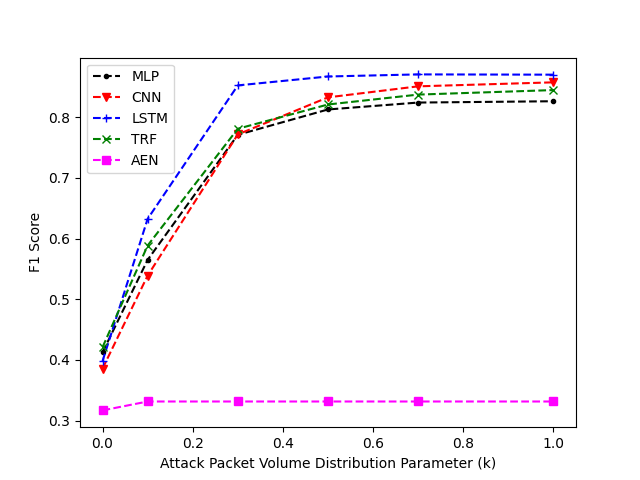}
                \caption{F1 Score}
                \label{fig:om_nc_f1_score}
            \end{subfigure}
            \begin{subfigure}[]{0.24\textwidth}
                \includegraphics[width=\textwidth]{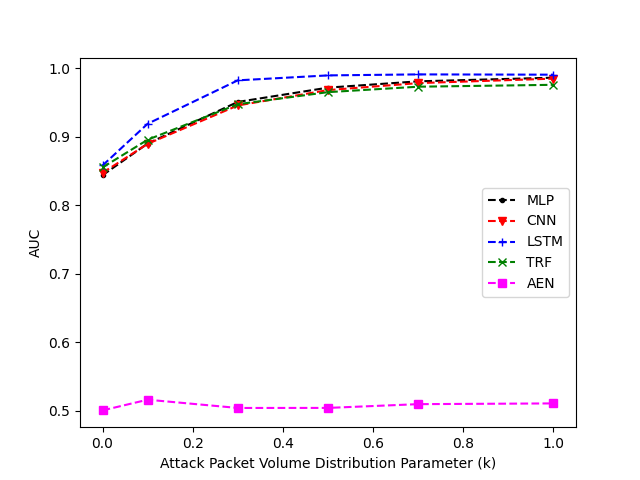}
                \caption{AUC}
                \label{fig:om_nc_auc}
            \end{subfigure}
            \begin{subfigure}[]{0.24\textwidth}
                \includegraphics[width=\textwidth]{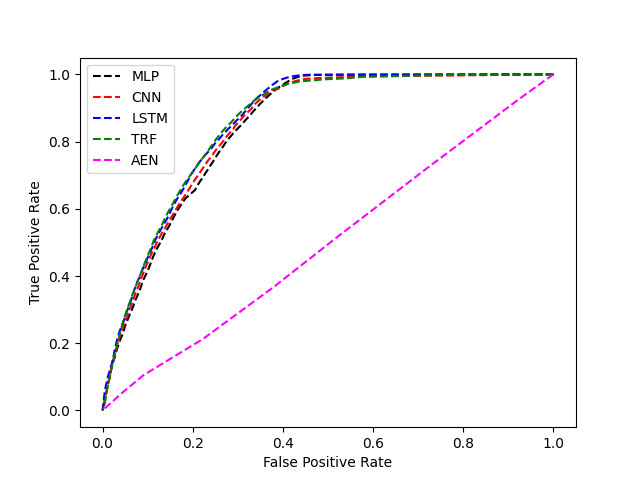}
                \caption{ROC Curve - $k = 0$}
                \label{fig:om_nc_roc_k_0}
            \end{subfigure}
            \caption{Compare different neural network models performance by using the one model without correlation (OM-NC) architecture}
            \label{fig:om_nc_performance}
        \end{figure}

        Figure \ref{fig:om_nc_performance} presents the performance of the one model without correlation architecture, i.e. OM-NC, across different neural network models. As we can see, the performance of all neural network models is increasing with higher values of $k$. We can again observe that LSTM neural network model has superior performance as compared to other models, with an F1 score between 0.4 and 0.86. Following that, CNN, TRF, and MLP show similar performance with an F1 score between 0.39 and 0.85. Again, AEN has the worst performance for the OM-NC architecture.

        The main take away from figures \ref{fig:mm_wc_performance}, \ref{fig:mm_nc_performance}, \ref{fig:om_wc_performance}, \ref{fig:om_nc_performance} are the following:
        \begin{itemize}
            \item Using the correlation information in MM-WC and OM-WC architectures for training the neural network models is helping them to predict the DDoS attacks with higher accuracy, especially for the lower values of $k$.
            \item LSTM is showing superior performance as compared to other neural network models in MM-WC, MM-NC, and OM-NC architectures. This is due to the fact that LSTM considers the temporal information embedded in the data much better as compared to other models, which helps it better detect the abnormal behavior of the nodes while they are under attack.
            \item For the OM-WC architecture, TRF is showing superior performance as compared to other neural network models. This supports the fact that the TRF model is also very good at considering the temporal information embedded in the data, and it is more suited to be used in a complex architecture like OM-WC with many input features.
            \item  In the architectures that we do not use the correlation information of the models, i.e. MM-NC and OM-NC, although we see worse performance for the models as compared to the architectures that use correlation information, the metrics of the models improve when we go from low values of $k$ to higher values of $k$. This is due to the fact that having a higher value for $k$ means the attackers are using higher packet volume for generating the DDoS attacks, which makes it easier for the neural network models to detect the attacks even without using the correlation information of other nodes' activity.
            \item The AEN model is performing worst as compared to all other models, almost in all scenarios. This is mainly attributed to the fact that for training the autoencoder model, we use the benign dataset traffic to learn the latent space attributed to the behavior of the nodes which are not under attack and then compare that latent space against the situations that nodes are under attack. Since we are training only on four days' worth of data, the autoencoder can not fully learn a meaningful latent space of the benign behavior of the nodes. The worst performance for the AEN model showed to be in the MM-WC architecture. In MM-WC architecture, we have very few samples since we are only using the samples of each IoT node individually as compared to OM-WC and OM-NC architectures, where we are using the samples of all nodes. Furthermore, MM-WC architecture also has considerably large input features due to using correlation information. Therefore, the AEN model has a hard time learning the behavior of the benign traffic and distinguishing it from the attack traffic. In this simulation, we used only 4 days' worth of data for training the models, but we expect that with increasing the number of training days, the AEN model should also perform better by taking in more samples of the benign traffic.
        \end{itemize}

        Since LSTM and TRF are performing better as compared to other models, here we also present the performance of only using either LSTM or TRF model against different architectures.
        
\begin{figure}[h]
            \centering 
            \begin{subfigure}[]{0.24\textwidth}
                \includegraphics[width=\textwidth]{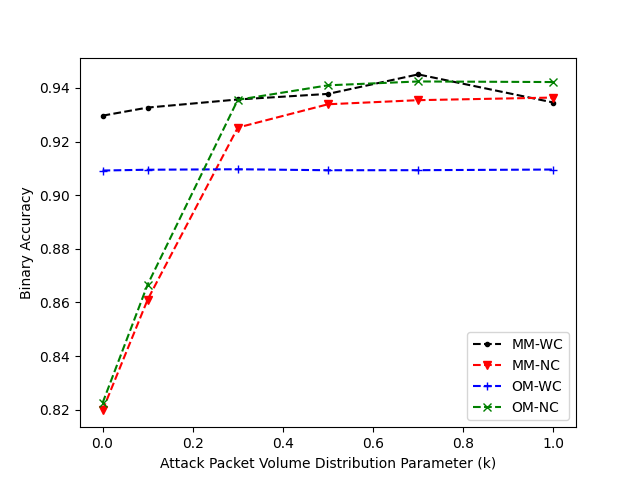}
                \caption{Binary Accuracy}
                \label{fig:lstm_binary_accuracy}
            \end{subfigure}
            \begin{subfigure}[]{0.24\textwidth}
                \includegraphics[width=\textwidth]{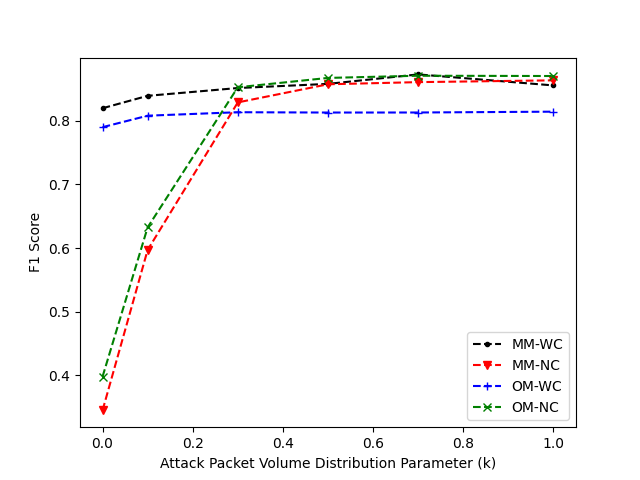}
                \caption{F1 Score}
                \label{fig:lstm_f1_score}
            \end{subfigure}
            \begin{subfigure}[]{0.24\textwidth}
                \includegraphics[width=\textwidth]{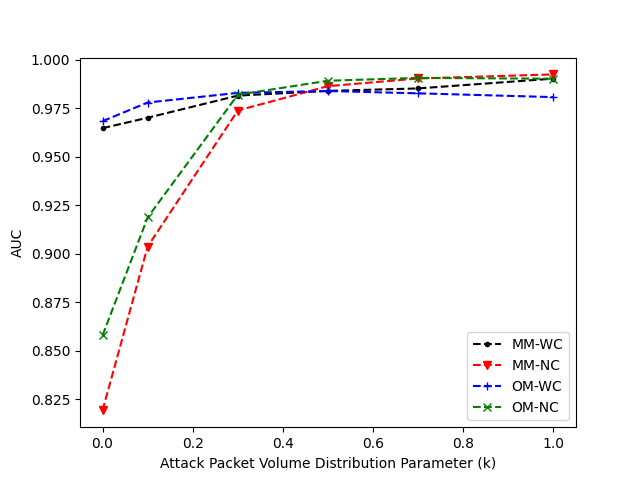}
                \caption{AUC}
                \label{fig:lstm_auc}
            \end{subfigure}
            \begin{subfigure}[]{0.24\textwidth}
                \includegraphics[width=\textwidth]{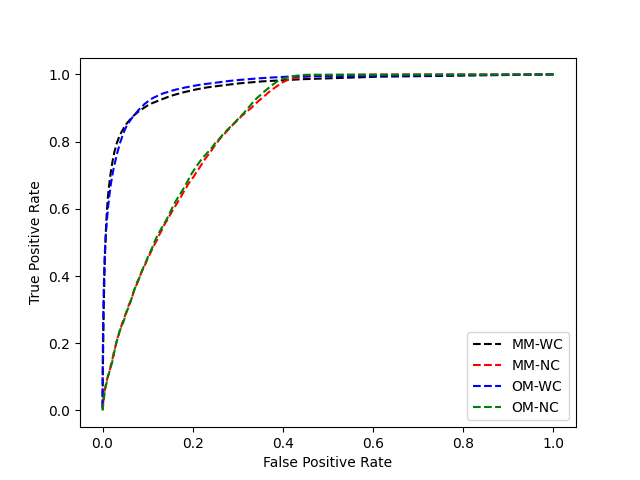}
                \caption{ROC Curve - $k = 0$}
                \label{fig:lstm_roc_k_0}
            \end{subfigure}
            \caption{Compare different architectures' performance by using the LSTM neural network model}
            \label{fig:lstm_performance}
        \end{figure}

        Figure \ref{fig:lstm_performance} presents the performance of the LSTM model across different architectures. As we can see, the architectures that do not use correlation information, i.e. MM-NC and OM-NC, show poor performance with an F1 score around 0.35 for low values of $k$. Their performance increase with higher values of $k$ with an F1 score around 0.85 which is similar to the performance of architectures that use correlation information, i.e. MM-WC and OM-WC. Furthermore, we can also observe that MM-WC architecture is performing the best in almost all values of $k$ as compared to other architectures with an F1 score between 0.82 and 0.87.
        
\begin{figure}[h]
            \centering 
            \begin{subfigure}[]{0.24\textwidth}
                \includegraphics[width=\textwidth]{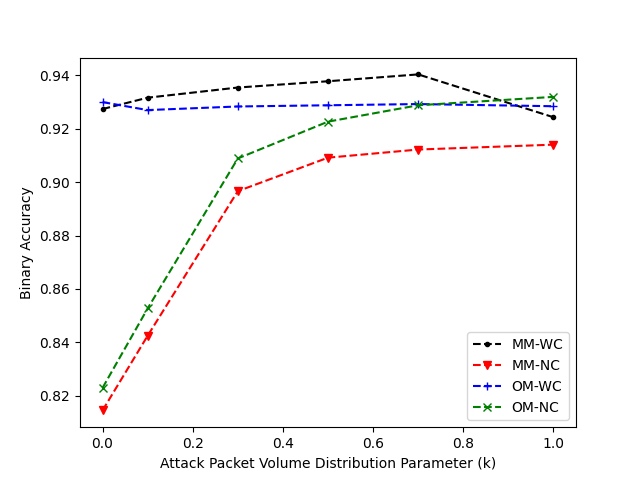}
                \caption{Binary Accuracy}
                \label{fig:trf_binary_accuracy}
            \end{subfigure}
            \begin{subfigure}[]{0.24\textwidth}
                \includegraphics[width=\textwidth]{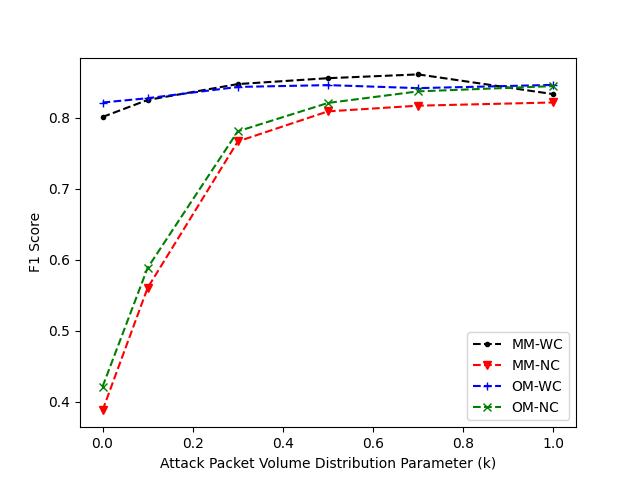}
                \caption{F1 Score}
                \label{fig:trf_f1_score}
            \end{subfigure}
            \begin{subfigure}[]{0.24\textwidth}
                \includegraphics[width=\textwidth]{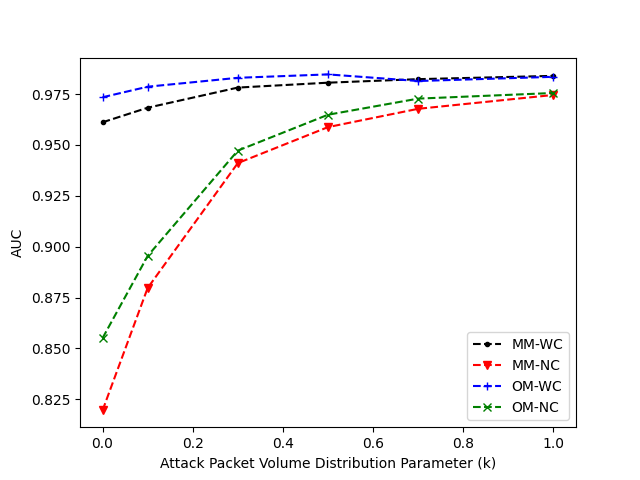}
                \caption{AUC}
                \label{fig:trf_auc}
            \end{subfigure}
            \begin{subfigure}[]{0.24\textwidth}
                \includegraphics[width=\textwidth]{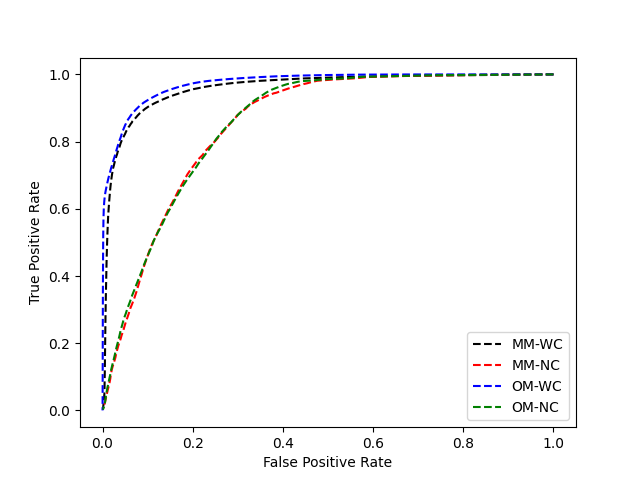}
                \caption{ROC Curve - $k = 0$}
                \label{fig:trf_roc_k_0}
            \end{subfigure}
            \caption{Compare different architectures' performance by using the TRF neural network model}
            \label{fig:trf_performance}
        \end{figure}

        Figure \ref{fig:trf_performance} presents the performance of the TRF model across different architectures. As we can see, again, the architectures that do not use correlation information, i.e. MM-NC and OM-NC, show poor performance with an F1 score around 0.4 for low values of $k$. Their performance increase with higher values of $k$ with an F1 score around 0.82 which is similar to the performance of architectures that use correlation information, i.e. MM-WC and OM-WC. Furthermore, we can also observe that both MM-WC and OM-WC architectures are showing similar good performance for all values of $k$ as compared to other architectures.

        All in all, we observe that using LSTM/MM-WC and TRF/OM-WC architecture provides the best and also similar performance. Depending on the application, one could use either of these models. In this paper, we prefer the LSTM/MM-WC architecture due to having distributed/individual models for each IoT node as compared to TRF/OM-WC architecture, where we only have one central model. Having a central model is not tolerant to failure, and the IoT system could hurt if that central model fails to do the DDoS detection. On the other hand, by having a distributed/individual model for each IoT node, the IoT system can survive even if some of the nodes fail to do the DDoS detection. In the rest of the paper, we will only consider the LSTM/MM-WC for further analysis.

    \subsection{DDoS Attack Prediction Analysis}
\begin{figure}[h!]
            \centering 
            \begin{subfigure}[]{0.24\textwidth}
                \includegraphics[width=\textwidth]{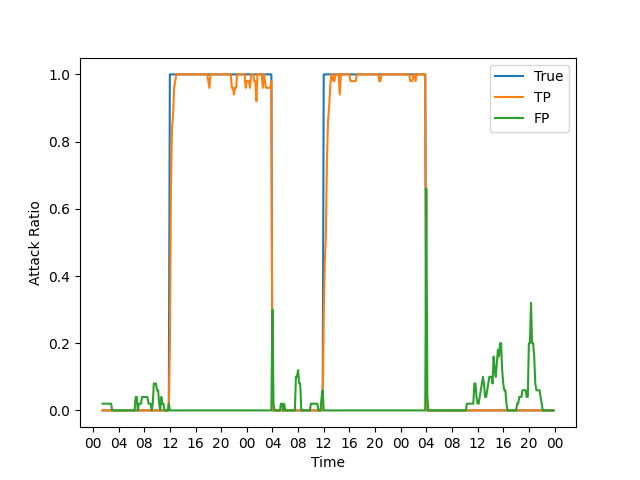}
                \caption{$k$ = 0}
                \label{fig:attack_prediction_k_0}
            \end{subfigure}
            \begin{subfigure}[]{0.24\textwidth}
                \includegraphics[width=\textwidth]{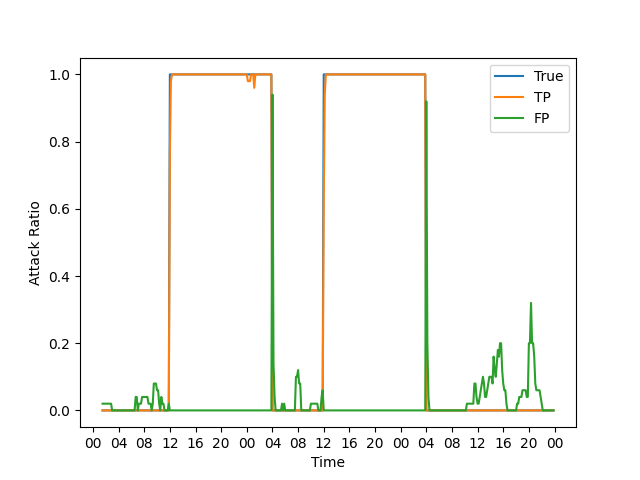}
                \caption{$k$ = 1}
                \label{fig:attack_prediction_k_1}
            \end{subfigure}
            \caption{Attack prediction vs time on the testing dataset for the case that attack starts at 12 pm for 8 hours over all IoT nodes.}
            \label{fig:attack_prediction}
        \end{figure}

        In this subsection, we present the DDoS attack prediction analysis by using the LSTM model and MM-WC architecture to see how fast and correctly in the real world our trained model could correctly predict the attacks. \\
        Figures \ref{fig:attack_prediction_k_0} and \ref{fig:attack_prediction_k_1} show the attack prediction for the case $k=0$ and $k=1$, respectively. The scenario that we selected for this figure is that the attacks start at 12 PM, which is the busiest time of the day when most of the nodes are active. The attack duration lasts for 16 hours. Each figure shows the ratio of the nodes that are under attack versus the time of the day. The ground truth for attacks (Labeled as True with the color blue) shows what ratio of the nodes are truly under attack. The model's prediction of true positive (labeled as TP with the color orange) shows the ratio of the nodes that are correctly predicted by the model to be under attack. The model's prediction false positive (labeled as FP with the color green) shows the ratio of the nodes that are mistakenly predicted by the model as being under attack. As we can see, for both $k=0$ and $k=1$, we are seeing the model is correctly predicting the nodes that are under attack with high accuracy. For the case of $k=1$, we are seeing a slightly better prediction of the attack while it is happening. This is due to the fact that for higher values of $k$, the attacker is sending higher packet volume as compared to lower values of $k$, and it would be easier for the model to detect the attack.
        
        Based on our analysis, the LSTM model using the MM-WC architecture could detect 100\% of the attack sessions described in section \ref{sec:attack-attack_dataset}. In fact, it could predict at least one of the IoT nodes which are truly under attack to have abnormal behavior. Furthermore, on average, 88\% of the IoT nodes that are truly under attack in all attack scenarios could correctly be predicted to have abnormal behavior. Additionally, on average, 92\% of the nodes which are truly not under attack could be truly predicted to have benign behavior. In terms of falsely predicting a node to be under attack, on average, the false positive detection lasts around 5.39 time slots. Furthermore, in terms of missing nodes that are truly under attack, on average, the false negative detection lasts around 4.41 time slots.
        
\begin{figure}
            \centering
            \includegraphics[width=\columnwidth]{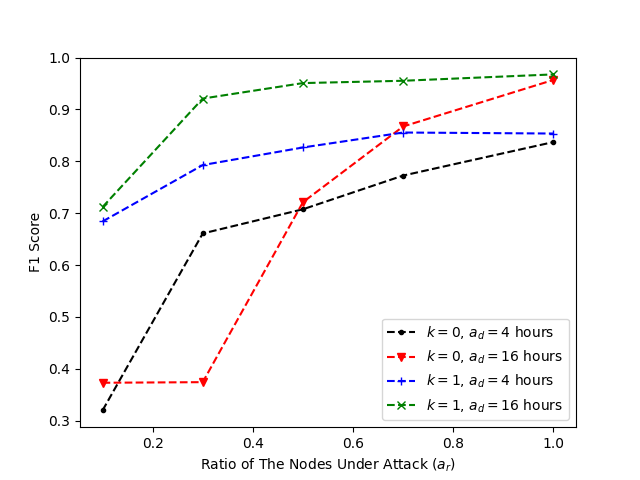}
            \caption{F1 Score vs ratio of the nodes under attack ($a_r$)}
            \label{fig:f1_vs_ar}
        \end{figure}

        To further analyze the performance of the proposed detection mechanism, we created a new testing dataset that varies the ratio of nodes that goes under attack from 0.1 to 1. Figure \ref{fig:f1_vs_ar} shows the F1 score of the LSTM model using MM-WC architecture versus the ratio of the nodes that are under attack, i.e. $a_r$. We have separated the different attack scenarios by considering the four cases of $k = 0/1$ and $a_d = 4/16 hours$. As we can see, the scenario of $k = 1$ and $a_d = 16 hours$ have the highest F1 score since the attacker is transferring many packets from the IoT nodes, and it is easier to get detected. On the other hand, when $k = 0$, for both $a_d = 4/16 hours$, we can see that the performance of the detection model is suffered due to the attacker camouflaging the attack traffic with benign traffic, especially when also the attacker is using a very low number of IoT devices, i.e. $a_r = 0.1$ to perform the attack. Note that, the case of using a low number of IoT devices to perform the attack, i.e. $a_r = 0.1$, and also a small attack packet volume $k = 0$ is not very realistic due to the fact that although the attacker is sending some extra packets to the victim server, it can not essentially disturb the behavior of the victim server. Note that we believe having low performance in detecting the DDoS attacks in such mentioned not realistic scenarios is not a good way of analyzing the performance of our proposed model. However, we provided this simulation result just for the comprehensiveness of this study. 
        
    \subsection{IoT Nodes with Constrained Resources}
        Based on the simulation results provided in section \ref{sec:results-binary_classification_metrics}, using the correlation information of the IoT nodes provides insightful information for the neural network models to make a much better prediction, especially for the case the attacker camouflages the attack in the benign traffic, i.e. $k=0$. As we mentioned in section \ref{sec:iot_constrained}, using the correlation information of \textit{all nodes} is infeasible in the real world, given the huge number of IoT devices that could go under attack. Therefore, we introduced five methods in section \ref{sec:iot_constrained} to compare different techniques for incorporating the correlation information of the IoT nodes that have the highest impact on correctly predicting the DDoS attacks. In this subsection, we present the performance of different methods by using the information of 5 nodes, including the node that is running the model's prediction. Furthermore, we only consider the LSTM neural network model and MM-WC architecture as our detection method.
        
\begin{figure}
            \centering
            \includegraphics[width=\columnwidth]{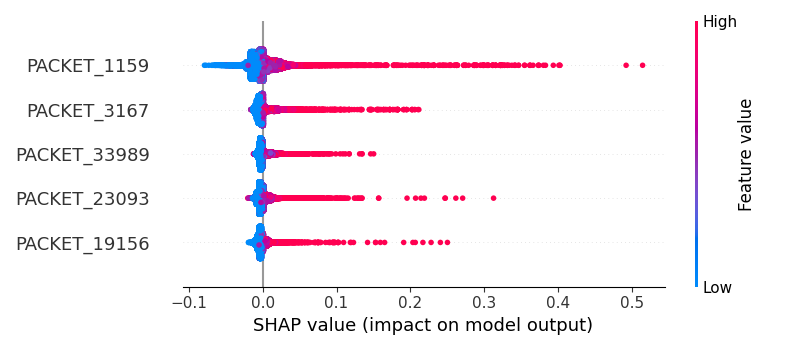}
            \caption{Feature importance analysis based on the LSTM/MM-WC model for node 1159}
            \label{fig:shap}
        \end{figure}

        One of the methods mentioned in section \ref{sec:iot_constrained} uses the most important features in the training dataset based on the SHAP values to figure out the top 5 most important nodes. Figure \ref{fig:shap} shows the top 5 important features in training the LSTM/MM-WC model for node 1159. The left vertical axis shows the feature names, and the features are sorted based on their importance in descending order. The right vertical axis shows the color heat map associated with the actual value of the feature, i.e. color shows whether the actual value of the feature is high (red color) or low (blue color) for each sample in the testing dataset. The horizontal axis shows the impact of each sample on the neural network model's prediction. In fact, higher(lower) values in the horizontal axis show that a sample has a higher(lower) prediction value. In our scenario, a higher prediction value shows an attack is happening, and a lower prediction value shows an attack is not happening. Figure \ref{fig:shap} supports the fact that having a high packet volume (feature value) for node 1159, has a high impact (SHAP value) on predicting a DDoS attack happening by using node 1159. Furthermore, using the packet volume of the other four nodes, namely 3167, 33989, 23093, and 19159 helped the neural network model trained by node 1159 to predict whether a DDoS attack was happening or not. Therefore, in the SHAP method discussed in section \ref{sec:iot_constrained}, in order to train the LSTM neural network based on the MM-WC architecture for node 1159, we will only use the correlation information from nodes 1159, 3167, 33989, 23093, and 19159. We repeat the same procedure for other nodes to determine the top 5 important features for training the LSTM/MM-WC model for each node.

\begin{figure}[h]
            \centering 
            \begin{subfigure}[]{0.24\textwidth}
                \includegraphics[width=\textwidth]{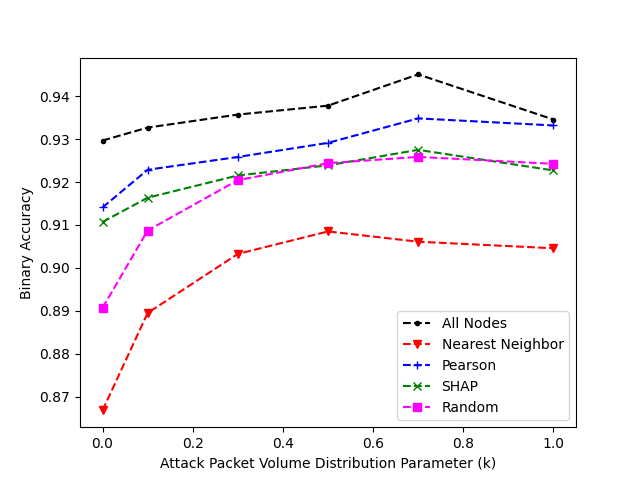}
                \caption{Binary Accuracy}
                \label{fig:meta_lstm_binary_accuracy}
            \end{subfigure}
            \begin{subfigure}[]{0.24\textwidth}
                \includegraphics[width=\textwidth]{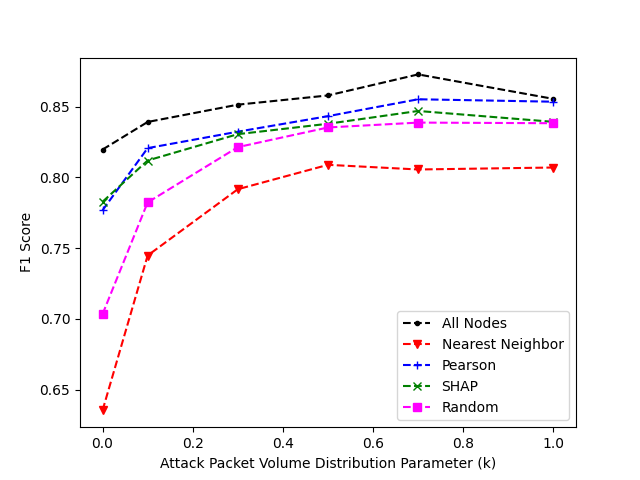}
                \caption{F1 Score}
                \label{fig:meta_lstm_f1_score}
            \end{subfigure}
            \begin{subfigure}[]{0.24\textwidth}
                \includegraphics[width=\textwidth]{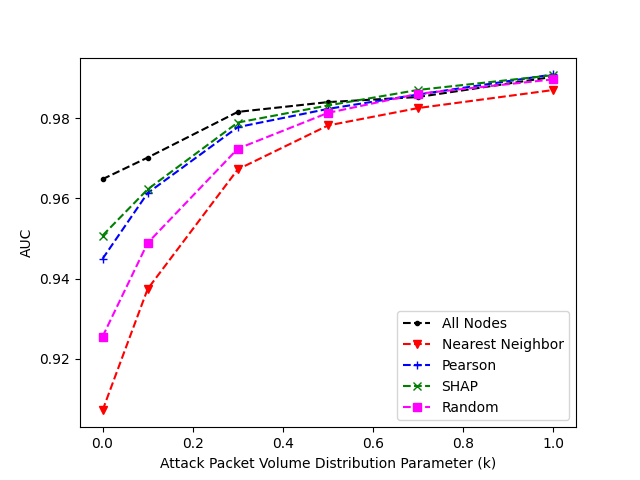}
                \caption{AUC}
                \label{fig:meta_lstm_auc}
            \end{subfigure}
            \begin{subfigure}[]{0.24\textwidth}
                \includegraphics[width=\textwidth]{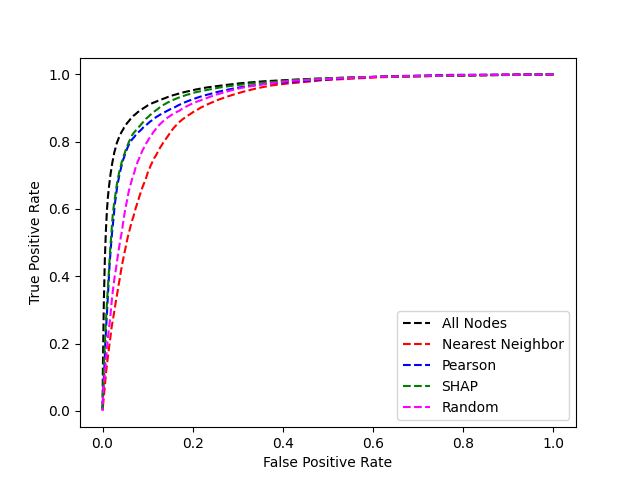}
                \caption{ROC Curve - $k = 0$}
                \label{fig:meta_lstm_roc_k_0}
            \end{subfigure}
            \caption{Compare the performance of different methods for selecting subset of the nodes for training the LSTM neural network model}
            \label{fig:meta_lstm_performance}
        \end{figure}

        Figure \ref{fig:meta_lstm_performance} compares the performance of the different techniques discussed in section \ref{sec:iot_constrained} by using the LSTM/MM-WC model and also considering the top 5 important nodes. Figure \ref{fig:meta_lstm_performance} supports the assumption that using the information of all nodes provides higher performance with an F1 score of between 0.82 and 0.87 as compared to other techniques, which only use the information of 5 nodes. Furthermore, we can see that using the Pearson correlation provides higher performance as compared to the SHAP, random, and nearest neighbor method with an F1 score degradation of up to 5\% as compared to using the information of all nodes. Using the SHAP technique to determine the most important features has a slightly lower performance as compared to the Pearson correlation method. Interestingly, the nearest neighbor method shows the worst performance of all. With the help of these results, we can conclude that by using the Pearson correlation, we can detect DDoS attacks over a large number of IoT devices by using the correlation information of only a fraction of IoT nodes with a slight degradation in performance. 
        
    \subsection{DDoS Detection Performance Analysis over All nodes In The Dataset}
\begin{figure}[h]
            \centering 
            \begin{subfigure}[]{0.24\textwidth}
                \includegraphics[width=\textwidth]{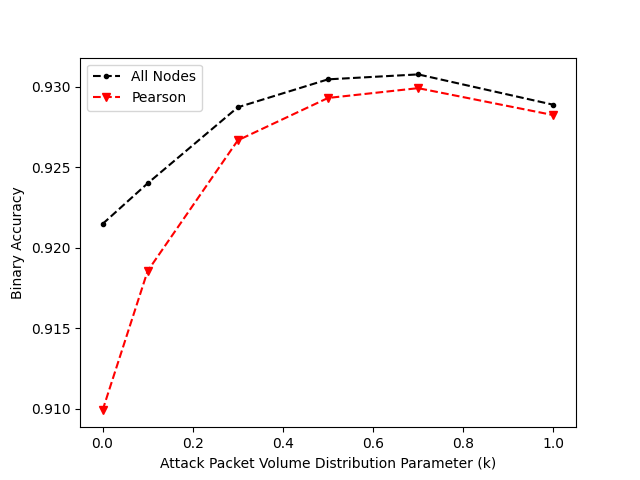}
                \caption{Binary Accuracy}
                \label{fig:all_nodes_binary_accuracy}
            \end{subfigure}
            \begin{subfigure}[]{0.24\textwidth}
                \includegraphics[width=\textwidth]{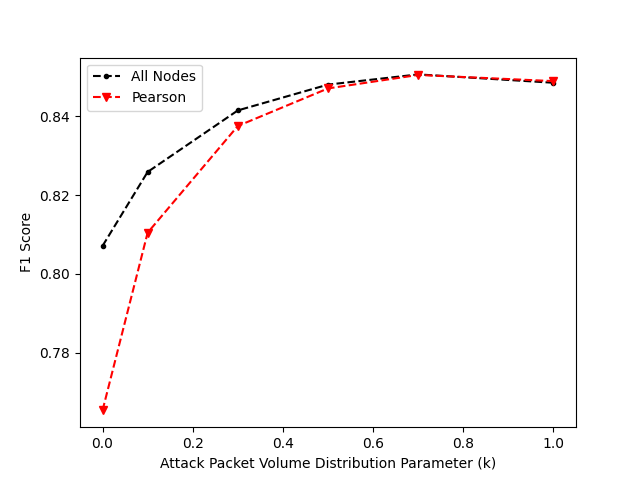}
                \caption{F1 Score}
                \label{fig:all_nodes_f1_score}
            \end{subfigure}
            \begin{subfigure}[]{0.24\textwidth}
                \includegraphics[width=\textwidth]{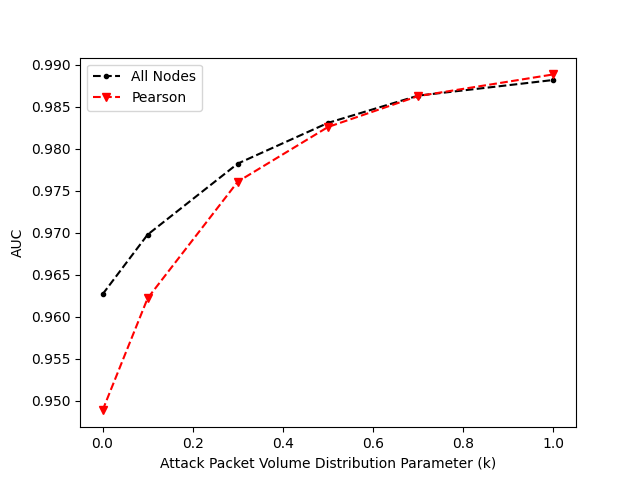}
                \caption{AUC}
                \label{fig:all_nodes_auc}
            \end{subfigure}
            \begin{subfigure}[]{0.24\textwidth}
                \includegraphics[width=\textwidth]{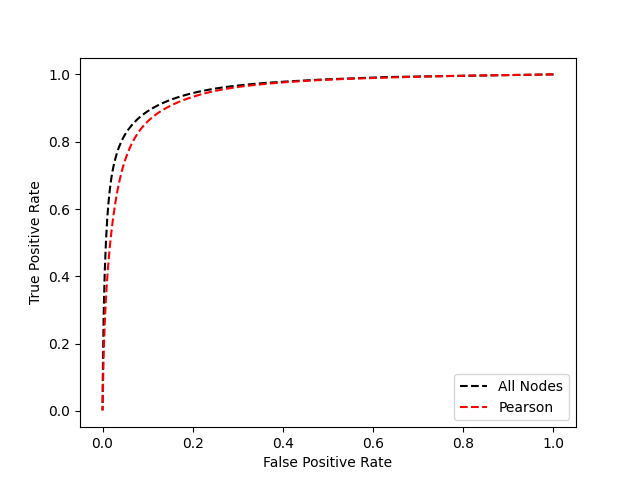}
                \caption{ROC Curve - $k = 0$}
                \label{fig:all_nodes_roc_k_0}
            \end{subfigure}
            \caption{Compare the performance of DDoS detection methods over all nodes}
            \label{fig:all_nodes_performance}
        \end{figure}

        In this subsection, we analyze the performance of the proposed DDoS detection mechanism over \textit{all} nodes of the dataset. For this purpose, we assume all nodes are randomly distributed into groups of 50 nodes, i.e., each group of 50 nodes can have access to the information of other nodes in that group. In total, we will have 81 groups of 50 IoT nodes. We use the LSTM neural network model and the MM-WC architecture. We also consider the \textit{All Nodes} and \textit{Pearson Correlation} methods for training and testing the neural network models that were introduced in section \ref{sec:iot_constrained}. Figure \ref{fig:all_nodes_performance} shows the average performance of the LSTM/MM-WC model by using both \textit{All Nodes} and \textit{Pearson Correlation} method over different attack packet distribution parameter, i.e. $k$. As we can see, if we use the correlation information of all 50 nodes in each group, we will have an F1 score between 0.81 and 0.85 while using the \textit{Pearson Correlation} method, we are observing up to 5\% lower performance in F1 score as compared to the \textit{All Nodes} method.

\section{Conclusion}
\label{sec:conclusion}    
    
    In this paper, we presented a machine-learning-based mechanism for detecting DDoS attacks in the ever-growing IoT systems. We used our previously published dataset that is based on real urban IoT activity data and emulated benign and attack traffic by using truncated Cauchy distribution. We introduced a new general training dataset that incorporated the correlation information of IoT nodes to be used by correlation-aware architectures for model training. Four new architectures, namely MM-WC, MM-NC, OM-WC, and OM-NC introduced that consider the case of either using a central or distributed neural network model and also consider either using the information of individual IoT nodes or using the correlation information of the IoT nodes. Furthermore, we studied the performance of five neural network models, namely MLP, CNN, LSTM, TRF, and AEN, by using the combination of all proposed architectures. Our extensive simulation results indicated that LSTM/MM-WC and TRF/OM-WC models have the best performance as compared to other combinations of architectures and neural network models. Furthermore, we observed that the architectures that use the correlation information of the nodes, i.e. MM-WC and OM-WC, have better performance as compared to the architectures that do not utilize the correlation information, especially in the case the attacker tries to camouflage the attack by mimicking the behavior of the benign traffic. Since the attacker could potentially use a massive number of IoT devices to perform a DDoS attack, it is not feasible to use the correlation information of all nodes for training neural network models. Therefore, we introduced heuristic solutions that actively select fewer number of IoT nodes for the purpose of training correlation-aware neural network models and detecting DDoS attacks. Our analysis indicated that selecting IoT nodes based on having the highest Pearson correlation for the purpose of training the correlation-aware architectures has the best performance as compared to other solutions. In future work, we plan to consider more complex detection architectures by using a collaborative method that uses IoT nodes, edge devices, and cloud servers. Furthermore, we want to explore the case of dynamically training and updating neural network models to make them capable of detecting DDoS attacks that use different behavior through time.

\section{Acknowledgments}

   This material is based upon work supported by Defense Advanced Research Projects Agency (DARPA) under Contract No. HR001120C0160 for the Open, Programmable, Secure 5G (OPS-5G) program. Any views, opinions, and/or findings expressed are those of the author(s) and should not be interpreted as representing the official views or policies of the Department of Defense or the U.S. Government.

\bibliographystyle{IEEEtran}
%\singlespacing
\bibliography{main.bib}

% Generated by IEEEtran.bst, version: 1.14 (2015/08/26)
\begin{thebibliography}{10}
\providecommand{\url}[1]{#1}
\csname url@samestyle\endcsname
\providecommand{\newblock}{\relax}
\providecommand{\bibinfo}[2]{#2}
\providecommand{\BIBentrySTDinterwordspacing}{\spaceskip=0pt\relax}
\providecommand{\BIBentryALTinterwordstretchfactor}{4}
\providecommand{\BIBentryALTinterwordspacing}{\spaceskip=\fontdimen2\font plus
\BIBentryALTinterwordstretchfactor\fontdimen3\font minus
  \fontdimen4\font\relax}
\providecommand{\BIBforeignlanguage}[2]{{%
\expandafter\ifx\csname l@#1\endcsname\relax
\typeout{** WARNING: IEEEtran.bst: No hyphenation pattern has been}%
\typeout{** loaded for the language `#1'. Using the pattern for}%
\typeout{** the default language instead.}%
\else
\language=\csname l@#1\endcsname
\fi
#2}}
\providecommand{\BIBdecl}{\relax}
\BIBdecl

\bibitem{shah2016survey}
S.~H. Shah and I.~Yaqoob, ``A survey: Internet of things ({IoT}) technologies,
  applications and challenges,'' in \emph{2016 IEEE Smart Energy Grid
  Engineering (SEGE)}.\hskip 1em plus 0.5em minus 0.4em\relax IEEE, 2016, pp.
  381--385.

\bibitem{7555867}
H.~Arasteh, V.~Hosseinnezhad, V.~Loia, A.~Tommasetti, O.~Troisi,
  M.~Shafie-khah, and P.~Siano, ``{IoT}-based smart cities: A survey,'' in
  \emph{2016 IEEE 16th International Conference on Environment and Electrical
  Engineering (EEEIC)}, 2016, pp. 1--6.

\bibitem{numIoTDevices}
``{Number of Internet of Things (IoT) connected devices worldwide from 2019 to
  2021, with forecasts from 2022 to 2030},''
  \url{https://www.statista.com/statistics/1183457/iot-connected-devices-worldwide/},
  accessed: 11-27-2022.

\bibitem{HPstudy1}
``{Internet of things research study - 2014 report},''
  \url{https://d-russia.ru/wp-content/uploads/2015/10/4AA5-4759ENW.pdf},
  accessed: 11-27-2022.

\bibitem{HPstudy2}
``{Internet of things research study - 2015 report},''
  \url{https://www.alain-bensoussan.com/wp-content/uploads/2017/08/34794474.pdf},
  accessed: 11-27-2022.

\bibitem{hassan2019current}
W.~H. Hassan \emph{et~al.}, ``Current research on internet of things ({IoT})
  security: A survey,'' \emph{Computer networks}, vol. 148, pp. 283--294, 2019.

\bibitem{neshenko2019demystifying}
N.~Neshenko, E.~Bou-Harb, J.~Crichigno, G.~Kaddoum, and N.~Ghani,
  ``Demystifying {IoT} security: An exhaustive survey on {IoT} vulnerabilities
  and a first empirical look on internet-scale {IoT} exploitations,''
  \emph{IEEE Communications Surveys \& Tutorials}, vol.~21, no.~3, pp.
  2702--2733, 2019.

\bibitem{hallman2017ioddos}
R.~Hallman, J.~Bryan, G.~Palavicini, J.~Divita, and J.~Romero-Mariona,
  ``{IoDDoS}-the internet of distributed denial of sevice attacks,'' in
  \emph{2nd international conference on internet of things, big data and
  security. SCITEPRESS}, 2017, pp. 47--58.

\bibitem{kolias2017mirai}
C.~Kolias, G.~Kambourakis, A.~Stavrou, and J.~Voas, ``{DDoS} in the {IoT}:
  Mirai and other botnets,'' \emph{Computer}, vol.~50, no.~7, pp. 80--84, 2017.

\bibitem{bertino2017botnets}
E.~Bertino and N.~Islam, ``Botnets and internet of things security,''
  \emph{Computer}, vol.~50, no.~2, pp. 76--79, 2017.

\bibitem{verma2013UDP}
K.~Verma, H.~Hasbullah, and A.~Kumar, ``An efficient defense method against udp
  spoofed flooding traffic of denial of service (dos) attacks in vanet,'' in
  \emph{2013 3rd IEEE International Advance Computing Conference (IACC)}, 2013,
  pp. 550--555.

\bibitem{schuba1997syn}
C.~Schuba, I.~Krsul, M.~Kuhn, E.~Spafford, A.~Sundaram, and D.~Zamboni,
  ``Analysis of a denial of service attack on tcp,'' in \emph{Proceedings. 1997
  IEEE Symposium on Security and Privacy (Cat. No.97CB36097)}, 1997, pp.
  208--223.

\bibitem{suresh2011ddos}
M.~Suresh and R.~Anitha, ``Evaluating machine learning algorithms for detecting
  ddos attacks,'' vol. 196, 07 2011, pp. 441--452.

\bibitem{Yungaicela2021slowrate}
N.~M. Yungaicela-Naula, C.~Vargas-Rosales, and J.~A. Perez-Diaz, ``Sdn-based
  architecture for transport and application layer ddos attack detection by
  using machine and deep learning,'' \emph{IEEE Access}, vol.~9, pp.
  108\,495--108\,512, 2021.

\bibitem{margolis2017miari}
J.~Margolis, T.~T. Oh, S.~Jadhav, Y.~H. Kim, and J.~N. Kim, ``An in-depth
  analysis of the mirai botnet,'' in \emph{2017 International Conference on
  Software Security and Assurance (ICSSA)}, 2017, pp. 6--12.

\bibitem{sinanovic2017analysis}
H.~Sinanovi{\'c} and S.~Mrdovic, ``Analysis of mirai malicious software,'' in
  \emph{2017 25th International Conference on Software, Telecommunications and
  Computer Networks (SoftCOM)}.\hskip 1em plus 0.5em minus 0.4em\relax IEEE,
  2017, pp. 1--5.

\bibitem{marzano2018botnet}
A.~Marzano, D.~Alexander, O.~Fonseca, E.~Fazzion, C.~Hoepers,
  K.~Steding-Jessen, M.~H. P.~C. Chaves, I.~Cunha, D.~Guedes, and W.~Meira,
  ``The evolution of bashlite and mirai iot botnets,'' in \emph{2018 IEEE
  Symposium on Computers and Communications (ISCC)}, 2018, pp.
  00\,813--00\,818.

\bibitem{kelley2018reaper}
T.~Kelley and E.~Furey, ``Getting prepared for the next botnet attack :
  Detecting algorithmically generated domains in botnet command and control,''
  in \emph{2018 29th Irish Signals and Systems Conference (ISSC)}, 2018, pp.
  1--6.

\bibitem{vishwakarma2020botnet}
\BIBentryALTinterwordspacing
R.~Vishwakarma and A.~K. Jain, ``A survey of ddos attacking techniques and
  defence mechanisms in the iot network,'' \emph{Telecommunication Systems},
  vol.~73, pp. 3--25, 1 2020. [Online]. Available:
  \url{https://link.springer.com/article/10.1007/s11235-019-00599-z}
\BIBentrySTDinterwordspacing

\bibitem{mcdermott2018mirai}
C.~D. McDermott, F.~Majdani, and A.~V. Petrovski, ``Botnet detection in the
  internet of things using deep learning approaches,'' in \emph{2018
  International Joint Conference on Neural Networks (IJCNN)}, 2018, pp. 1--8.

\bibitem{sharma2016machine}
N.~Sharma, A.~Mahajan, and V.~Malhotra, ``Machine learning techniques used in
  detection of {DoS} attacks: a literature review,'' \emph{International
  Journal of Advance Research in Computer Science and Software Engineering},
  vol.~6, no.~3, pp. 100--105, 2016.

\bibitem{meidan2018mirai}
Y.~Meidan, M.~Bohadana, Y.~Mathov, Y.~Mirsky, A.~Shabtai, D.~Breitenbacher, and
  Y.~Elovici, ``N-baiot—network-based detection of {IoT} botnet attacks using
  deep autoencoders,'' \emph{IEEE Pervasive Computing}, vol.~17, no.~3, pp.
  12--22, 2018.

\bibitem{CICIDS2017}
I.~Sharafaldin, A.~Habibi~Lashkari, and A.~Ghorbani, ``Toward generating a new
  intrusion detection dataset and intrusion traffic characterization,'' in
  \emph{4th International Conference on Information Systems Security and
  Privacy (ICISSP)}, 01 2018, pp. 108--116.

\bibitem{hekmati2021workshop}
\BIBentryALTinterwordspacing
A.~Hekmati, E.~Grippo, and B.~Krishnamachari, ``Large-scale urban iot activity
  data for ddos attack emulation,'' in \emph{Proceedings of the 19th ACM
  Conference on Embedded Networked Sensor Systems}, ser. SenSys '21.\hskip 1em
  plus 0.5em minus 0.4em\relax New York, NY, USA: Association for Computing
  Machinery, 2021, p. 560–564. [Online]. Available:
  \url{https://doi.org/10.1145/3485730.3493695}
\BIBentrySTDinterwordspacing

\bibitem{hekmati2022conference}
------, ``Neural networks for ddos attack detection using an enhanced urban iot
  dataset,'' in \emph{2022 International Conference on Computer Communications
  and Networks (ICCCN)}, 2022, pp. 1--8.

\bibitem{field2002network}
T.~Field, U.~Harder, and P.~Harrison, ``Network traffic behaviour in switched
  ethernet systems,'' in \emph{Proceedings. 10th IEEE International Symposium
  on Modeling, Analysis and Simulation of Computer and Telecommunications
  Systems}.\hskip 1em plus 0.5em minus 0.4em\relax IEEE, 2002, pp. 33--42.

\bibitem{doshi2018ddosML}
R.~Doshi, N.~Apthorpe, and N.~Feamster, ``Machine learning ddos detection for
  consumer internet of things devices,'' in \emph{2018 IEEE Security and
  Privacy Workshops (SPW)}, 2018, pp. 29--35.

\bibitem{chen2020sdnML}
Y.-W. Chen, J.-P. Sheu, Y.-C. Kuo, and N.~Van~Cuong, ``Design and
  implementation of iot ddos attacks detection system based on machine
  learning,'' in \emph{2020 European Conference on Networks and Communications
  (EuCNC)}, 2020, pp. 122--127.

\bibitem{mohammed2018sdnML}
S.~S. Mohammed, R.~Hussain, O.~Senko, B.~Bimaganbetov, J.~Lee, F.~Hussain,
  C.~A. Kerrache, E.~Barka, and M.~Z. Alam~Bhuiyan, ``A new machine
  learning-based collaborative ddos mitigation mechanism in software-defined
  network,'' in \emph{2018 14th International Conference on Wireless and Mobile
  Computing, Networking and Communications (WiMob)}, 2018, pp. 1--8.

\bibitem{syed2020brokerML}
\BIBentryALTinterwordspacing
N.~F. Syed, Z.~Baig, A.~Ibrahim, and C.~Valli, ``Denial of service attack
  detection through machine learning for the iot,'' \emph{Journal of
  Information and Telecommunication}, vol.~4, no.~4, pp. 482--503, 2020.
  [Online]. Available: \url{https://doi.org/10.1080/24751839.2020.1767484}
\BIBentrySTDinterwordspacing

\bibitem{UNB_two}
``{University of New Brunswick, “DDoS Evaluation Dataset
  (CICDDoS2019)”,unb.ca, 2019},''
  \url{https://www.unb.ca/cic/datasets/ddos-2019.html}, accessed: 09-13-2021.

\bibitem{Nugraha2020slowrate}
B.~Nugraha and R.~N. Murthy, ``Deep learning-based slow ddos attack detection
  in sdn-based networks,'' in \emph{2020 IEEE Conference on Network Function
  Virtualization and Software Defined Networks (NFV-SDN)}, 2020, pp. 51--56.

\bibitem{Muraleedharan2021slowrate}
\BIBentryALTinterwordspacing
N.~Muraleedharan and B.~Janet, ``Flow-based machine learning approach for slow
  http distributed denial of service attack classification,''
  \emph{International Journal of Computational Science and Engineering},
  vol.~24, no.~2, pp. 147--161, 2021. [Online]. Available:
  \url{https://www.inderscienceonline.com/doi/abs/10.1504/IJCSE.2021.115101}
\BIBentrySTDinterwordspacing

\bibitem{Cheng2020slowrate}
\BIBentryALTinterwordspacing
H.~Cheng, J.~Liu, T.~Xu, B.~Ren, J.~Mao, and W.~Zhang, ``Machine learning based
  low-rate ddos attack detection for sdn enabled iot networks,''
  \emph{International Journal of Sensor Networks}, vol.~34, no.~1, pp. 56--69,
  2020. [Online]. Available:
  \url{https://www.inderscienceonline.com/doi/abs/10.1504/IJSNET.2020.109720}
\BIBentrySTDinterwordspacing

\bibitem{liu2018sdnML}
Y.~Liu, M.~Dong, K.~Ota, J.~Li, and J.~Wu, ``Deep reinforcement learning based
  smart mitigation of ddos flooding in software-defined networks,'' in
  \emph{2018 IEEE 23rd International Workshop on Computer Aided Modeling and
  Design of Communication Links and Networks (CAMAD)}, 2018, pp. 1--6.

\bibitem{roopak2019mlIoT}
M.~Roopak, G.~Yun~Tian, and J.~Chambers, ``Deep learning models for cyber
  security in iot networks,'' in \emph{2019 IEEE 9th Annual Computing and
  Communication Workshop and Conference (CCWC)}, 2019, pp. 0452--0457.

\bibitem{gurulakshmi2018mlIoT}
K.~Gurulakshmi and A.~Nesarani, ``Analysis of iot bots against ddos attack
  using machine learning algorithm,'' in \emph{2018 2nd International
  Conference on Trends in Electronics and Informatics (ICOEI)}, 2018, pp.
  1052--1057.

\bibitem{nslkddDataset}
``{NSL-KDD dataset},'' \url{https://www.unb.ca/cic/datasets/nsl.html},
  accessed: 11-27-2022.

\bibitem{zekri2017mlcloud}
M.~Zekri, S.~E. Kafhali, N.~Aboutabit, and Y.~Saadi, ``Ddos attack detection
  using machine learning techniques in cloud computing environments,'' in
  \emph{2017 3rd International Conference of Cloud Computing Technologies and
  Applications (CloudTech)}, 2017, pp. 1--7.

\bibitem{Blaise2020MLIoT}
A.~Blaise, M.~Bouet, V.~Conan, and S.~Secci, ``Botnet fingerprinting: A
  frequency distributions scheme for lightweight bot detection,'' \emph{IEEE
  Transactions on Network and Service Management}, vol.~17, no.~3, pp.
  1701--1714, 2020.

\bibitem{ctu13dataset}
``{The CTU-13 Dataset. A Labeled Dataset with Botnet, Normal and Background
  traffic.}'' \url{https://www.stratosphereips.org/datasets-ctu13}, accessed:
  11-27-2022.

\bibitem{Soe2020MLIoT}
\BIBentryALTinterwordspacing
Y.~N. Soe, Y.~Feng, P.~I. Santosa, R.~Hartanto, and K.~Sakurai, ``Machine
  learning-based iot-botnet attack detection with sequential architecture,''
  \emph{Sensors}, vol.~20, no.~16, 2020. [Online]. Available:
  \url{https://www.mdpi.com/1424-8220/20/16/4372}
\BIBentrySTDinterwordspacing

\bibitem{Nugraha2020MLIoT}
B.~Nugraha, A.~Nambiar, and T.~Bauschert, ``Performance evaluation of botnet
  detection using deep learning techniques,'' in \emph{2020 11th International
  Conference on Network of the Future (NoF)}, 2020, pp. 141--149.

\bibitem{Kumar2019MLIoT}
A.~Kumar and T.~J. Lim, ``Edima: Early detection of iot malware network
  activity using machine learning techniques,'' in \emph{2019 IEEE 5th World
  Forum on Internet of Things (WF-IoT)}, 2019, pp. 289--294.

\bibitem{DARPA_2000}
``{DARPA 2000 Intrustion Detection Scenario Specific Data Sets},''
  \url{https://www.ll.mit.edu/r-d/datasets/}, accessed: 10-08-2021.

\bibitem{CAIDA_UCSD}
``{The CAIDA UCSD ”DDoS Attack 2007” Dataset, 2007},''
  \url{https://www.caida.org/catalog/datasets/ddos-20070804_dataset/},
  accessed: 10-08-2021.

\bibitem{shiravi2012toward}
A.~Shiravi, H.~Shiravi, M.~Tavallaee, and A.~A. Ghorbani, ``Toward developing a
  systematic approach to generate benchmark datasets for intrusion detection,''
  \emph{computers \& security}, vol.~31, no.~3, pp. 357--374, 2012.

\bibitem{UNB_three}
``{University of New Brunswick, “CSE-CIC-IDS2018 on AWS”, 2018},''
  \url{https://www.unb.ca/cic/datasets/ids-2018.html}, accessed: 09-13-2021.

\bibitem{UNSW_two}
``{University of New South Wales, The Bot-IoT Dataset},''
  \url{https://research.unsw.edu.au/projects/bot-iot-dataset}, accessed:
  09-12-2021.

\bibitem{bot_iot}
``{A Scheme for Generating a Dataset for Anomalous Activity Detection in {IoT}
  Networks},''
  \url{https://sites.google.com/view/iot-network-intrusion-dataset}, accessed:
  09-12-2021.

\bibitem{erhan2020bougazicci}
D.~Erhan and E.~Anar{\i}m, ``Bo{\u{g}}azi{\c{c}}i {University} distributed
  denial of service dataset,'' \emph{Data in brief}, vol.~32, p. 106187, 2020.

\bibitem{chandrasekaran2009survey}
B.~Chandrasekaran, ``Survey of network traffic models,'' \emph{Washington
  University in St. Louis CSE}, vol. 567, 2009.

\bibitem{Pal1992MLP}
S.~Pal and S.~Mitra, ``Multilayer perceptron, fuzzy sets, and classification,''
  \emph{IEEE Transactions on Neural Networks}, vol.~3, no.~5, pp. 683--697,
  1992.

\bibitem{Albawi2017CNN}
S.~Albawi, T.~A. Mohammed, and S.~Al-Zawi, ``Understanding of a convolutional
  neural network,'' in \emph{2017 International Conference on Engineering and
  Technology (ICET)}, 2017, pp. 1--6.

\bibitem{Hochreiter1997LSTM}
S.~Hochreiter and J.~Schmidhuber, ``Long short-term memory,'' \emph{Neural
  computation}, vol.~9, pp. 1735--80, 12 1997.

\bibitem{Vaswani2017TRF}
\BIBentryALTinterwordspacing
A.~Vaswani, N.~Shazeer, N.~Parmar, J.~Uszkoreit, L.~Jones, A.~N. Gomez, L.~u.
  Kaiser, and I.~Polosukhin, ``Attention is all you need,'' in \emph{Advances
  in Neural Information Processing Systems}, I.~Guyon, U.~V. Luxburg,
  S.~Bengio, H.~Wallach, R.~Fergus, S.~Vishwanathan, and R.~Garnett, Eds.,
  vol.~30.\hskip 1em plus 0.5em minus 0.4em\relax Curran Associates, Inc.,
  2017. [Online]. Available:
  \url{https://proceedings.neurips.cc/paper/2017/file/3f5ee243547dee91fbd053c1c4a845aa-Paper.pdf}
\BIBentrySTDinterwordspacing

\bibitem{Bank2020AEN}
\BIBentryALTinterwordspacing
D.~Bank, N.~Koenigstein, and R.~Giryes, ``Autoencoders,'' 2020. [Online].
  Available: \url{https://arxiv.org/abs/2003.05991}
\BIBentrySTDinterwordspacing

\bibitem{tensorflow_unbalanced_dataset}
``{Classification on imbalanced data},''
  \url{https://www.tensorflow.org/tutorials/structured_data/imbalanced_data},
  accessed: 11-27-2022.

\bibitem{Lundberg2017SHAP}
\BIBentryALTinterwordspacing
S.~M. Lundberg and S.-I. Lee, ``A unified approach to interpreting model
  predictions,'' in \emph{Advances in Neural Information Processing Systems},
  I.~Guyon, U.~V. Luxburg, S.~Bengio, H.~Wallach, R.~Fergus, S.~Vishwanathan,
  and R.~Garnett, Eds., vol.~30.\hskip 1em plus 0.5em minus 0.4em\relax Curran
  Associates, Inc., 2017. [Online]. Available:
  \url{https://proceedings.neurips.cc/paper/2017/file/8a20a8621978632d76c43dfd28b67767-Paper.pdf}
\BIBentrySTDinterwordspacing

\end{thebibliography}

\end{document}